%% file: template.tex
\documentclass{vgtc}                          % final (conference style)
\graphicspath{{figures/}{pictures/}{images/}{./}} % where to search for the images

\usepackage{times}                     % we use Times as the main font
\usepackage{tabu}                      % only used for the table example
\usepackage{booktabs}                  % only used for the table example
\usepackage{lipsum}                    % used to generate placeholder text
\usepackage{mwe}                       % used to generate placeholder figures

\usepackage{mathptmx}                  % use matching math font
\usepackage{verbatim}
\usepackage{xspace}
\usepackage{enumitem}
\usepackage{booktabs}
\usepackage{bm}
\usepackage{multirow}
\usepackage[table]{xcolor}
\usepackage[cjk]{kotex}

\def\react{\textit{React}\xspace}
\def\hook{\texttt{Hook}\xspace}
\def\hooks{\texttt{Hooks}\xspace}
\def\hooklens{\textsc{HookLens}\xspace}
\def\useState{\texttt{useState}\xspace}
\def\useEffect{\texttt{useEffect}\xspace}
\def\component{\texttt{Component}\xspace}
\def\components{\texttt{Components}\xspace}
\def\prop{\texttt{Prop}\xspace}
\def\props{\texttt{Props}\xspace}
\def\state{\texttt{State}\xspace}
\def\states{\texttt{States}\xspace}
\def\effect{\texttt{Effect}\xspace}
\def\effects{\texttt{Effects}\xspace}

\usepackage{soul}
\usepackage{xcolor}
\usepackage{tikz}

\newcommand{\revised}[1]{\textcolor{black}{#1}}

\definecolor{statebg}{RGB}{205, 239, 179}
\definecolor{componentbg}{RGB}{226, 226, 226}
\definecolor{effectbg}{RGB}{187, 228, 255}
\definecolor{propbg}{RGB}{238, 208, 171}
\definecolor{exampletxt}{RGB}{50, 50, 50}

\sethlcolor{examplebg}

\onlineid{7611}

\vgtccategory{Research}

\vgtcinsertpkg

\title{HookLens: Visual Analytics for Understanding React Hooks Structures}
\author{
Suyeon Hwang\thanks{e-mail: stom.hwang@\{hcil.snu.ac.kr, samsung.com\}} $\text{ }^{1, 2}$ \quad
Minkyu Kweon\thanks{e-mail: \{mk, jmrhee, shlee, shpark, swjung, hj\}@hcil.snu.ac.kr} $\text{ }^{1}$ \quad  
Jeongmin Rhee\footnotemark[2] $\text{ }^{1}$ \quad 
Soohyun Lee\footnotemark[2] $\text{ }^{1}$ \quad\\
Seokhyeon Park\footnotemark[2] $\text{ }^{1}$ \quad
Seokweon Jung\footnotemark[2] $\text{ }^{1}$ \quad 
Hyeon Jeon\footnotemark[2] $\text{ }^{1}$ \quad
Jinwook Seo\thanks{e-mail: jseo@snu.ac.kr, corresponding author} $\text{ }^{1}$  \\ 
\scriptsize $^1$Seoul National University \quad 
\scriptsize $^2$Samsung Electronics \\ 
}

\teaser{
  \centering
  \includegraphics[width=\linewidth]{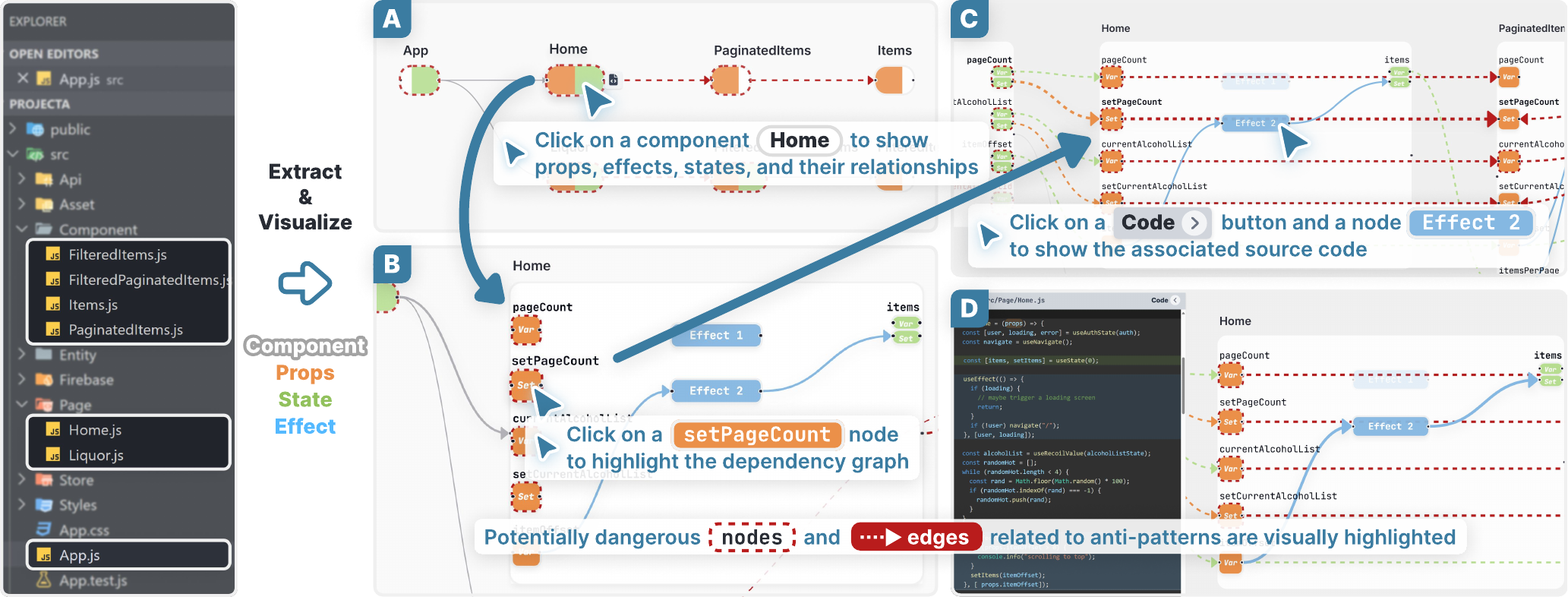}
  \caption{The overview of \hooklens. The nested directed node-link diagram visualizes the structure of a \react application. Nodes represent \components, \props, and \hooks---key elements for understanding \react---and edges indicate their relationships. Nodes for \props and \hooks are nested within \component nodes and are initially hidden. Users can click a \component node to reveal its internal \props, \hooks and their relationships, and then examine these details (A). Users can also click a \prop or \hook node to highlight its relevant nodes and edges (B). Potentially problematic \components, \props, \hooks, and relationships are highlighted with red outlines and edges (C). Moreover, users can view the associated source code in the code viewer (D).}
  \label{fig:teaser}
}

\abstract{
    \input{REVISION/0_abstract}
} % end of abstract


\begin{document}


\firstsection{Introduction}

\maketitle

%% \section{Introduction} %for journal use above \firstsection{..} instead
\input{REVISION/1_introduction}

\input{REVISION/2_background}

\input{REVISION/3_related_works}

\input{REVISION/4_tasks_and_design_requirements}

\input{REVISION/5_hooklens}

\input{REVISION/6_evaluation}

\input{REVISION/7_evalute_LLM_assistant}

\input{REVISION/8_discussion}

\input{REVISION/9_conclusion}

% \section*{Supplemental Materials}
% \label{sec:supplemental_materials}

% Refer to the instructions for this section (\cref{sec:supplement_inst}).
% Below is an example you can follow that includes the actual supplemental material for this template:

% All supplemental materials are available on OSF at \url{https://doi.org/10.17605/OSF.IO/2NBSG}, released under a CC BY 4.0 license.
% In particular, they include (1) Excel files containing the data for and analyses for creating \cref{tab:vis_papers} and \cref{fig:vis_papers}, (2) figure images in multiple formats, and (3) a full version of this paper with all appendices.
% Our other code is intellectual property of a corporation---Starbucks Research---and there is no feasible way to share it publicly.

% \section*{Figure Credits}
% \label{sec:figure_credits}

% Refer to the instructions for this section (\cref{sec:figure_credits_inst}).
% Here are the actual figure credits for this template:

% \Cref{fig:teaser} image credit: Scott Miller / Special to the Vancouver Sun, January 22, 2009, page A6.

% \Cref{fig:vis_papers} is a partial recreation of Fig.\ 1 from \cite{Isenberg:2017:VMC}, which is in the public domain.

% %% if specified like this the section will be committed in review mode
\acknowledgments{
This work was supported by the National Research Foundation of Korea (NRF) grant funded by the Korean government (MSIT) (No. 2023R1A2C200520911), the Institute of Information \& communications Technology Planning \& Evaluation (IITP) grant funded by the Korean government (MSIT) [NO.RS-2021-II211343, Artificial Intelligence Graduate School Program (Seoul National University)], and in part by Samsung Electronics. The ICT at Seoul National University provided research facilities for this study. Hyeon Jeon is in part supported by Google Ph.D. Fellowship.}
\vspace{-0.2cm}

\bibliographystyle{abbrv-doi}

\bibliography{template}
\end{document}

%% file: REVISION/0_abstract.tex
%
%
Maintaining and refactoring \react web applications is challenging, as \react code often becomes complex due to its core API called \hooks.
For example, \hooks often lead developers to create complex dependencies among components, making code behavior unpredictable and reducing maintainability, i.e., anti-patterns. 
To address this challenge, we present \hooklens, an interactive visual analytics system that helps developers understand how \hooks define dependencies and data flows between components.
Informed by an iterative design process with experienced \react developers, \hooklens supports users to efficiently understand the structure and dependencies between components and to identify anti-patterns.
A quantitative user study with 12 \react developers demonstrates that \hooklens significantly improves participants' accuracy in detecting anti-patterns compared to conventional code editors.
Moreover, a comparative study with state-of-the-art LLM-based coding assistants confirms that these improvements even surpass the capabilities of such coding assistants on the same task.

%% file: REVISION/1_introduction.tex
% 
% 
% \section{Introduction}
\label{sec:introduction}

\react\footnote{\url{https://react.dev/}} is widely used in modern web development, but maintaining and refactoring \react applications remains challenging.
Unlike conventional web development, where the code that defines the structure, functionality, and design of user interface (UI) components is scattered across multiple files, \react enables developers to implement these components as self-contained functions.
By doing so, \react helps developers build applications more intuitively and efficiently.
Furthermore, \hooks, the core API of \react, extend the functionality of each component, enabling the development of more dynamic and interactive applications.
However, \hooks often introduce complex dependencies among components, creating anti-patterns that make code unpredictable, difficult to understand, and challenging to maintain~\cite{FERREIRAPropDrilling, FardJavaScriptCodeSmells}.
For example, when \hooks are used across hierarchically nested components, developers frequently pass data through intermediate components that do not directly use it; this anti-pattern forces them to review multiple components when modifying a single one.

To address this challenge, we propose \hooklens, an interactive visual analytics system that supports developers in understanding the structure and dependencies of \react code from the perspective of \hooks.
We first conduct preliminary interviews with three experienced \react developers to identify the core tasks involved in understanding and refactoring \react applications.
We then conduct an iterative design process with eight additional experienced \react developers, enhancing the usability of \hooklens and its effectiveness in visually exploring code structure and identifying anti-patterns.
Our final prototype of \hooklens enables users to analyze the structure of applications, from component hierarchies to code-level details, through details-on-demand interactions.

To demonstrate the effectiveness and usability of \hooklens, we conduct a quantitative user study with 12 \react developers of varying levels of expertise.
Participants are asked to identify anti-patterns associated with \hooks in \react code using both \hooklens and a conventional code editor (\textit{VS Code}).
We further compare the result obtained from participants using \hooklens with those produced by state-of-the-art LLM-based coding assistants, such as \textit{Claude Code}, on the same task to examine the analytical capabilities of these assistants.
The results indicate that \hooklens substantially improves the accuracy in identifying anti-patterns compared to \textit{VS Code}, and even outperforms LLM-based assistants, underscoring the continued importance of visual analytics in understanding complex \react applications.

In summary, our contributions are as follows: 
\begin{itemize}[leftmargin=*, noitemsep, topsep=4pt]
    \item We identify core tasks and design requirements to understand \react applications based on interviews and an iterative design process.
    \item We introduce \hooklens, an interactive visual analytics system that helps developers better understand the code structure of \react applications and more effectively detect anti-patterns.
    \item We demonstrate the effectiveness and usability of \hooklens through a user study with real-world \react projects and a comparative study with LLM-based coding assistants.
\end{itemize}

\hooklens is available at \url{https://hook-lens.github.io/hook-lens/}, where users can freely upload their codebase or connect their GitHub repository.

%% file: REVISION/2_background.tex
\section{Background}
\label{sec:backgrounds}

We discuss the preliminaries essential for understanding the remainder of this paper.

\subsection{React}
\react is one of the most widely used libraries for developing web applications~\cite{soPopularTech}.
\react creates and manages individual UI elements as functions called \components.
The way each \component manages its data and functionalities is defined by a core API in \react called \hooks.

\vspace{4pt}
\noindent
\textbf{Components.}
\components are functions that define UI elements~\cite{FERREIRAMigrateClassToFunction, reactcomponent}.
\components can declare child \components and pass data to them via arguments called \props.
By tracing these hierarchies of \components, developers can easily understand the application's UI and behavior.

\vspace{4pt}
\noindent
\textbf{Hooks.}
\label{sec:hooks}
\hooks allow \react applications to perform complex operations by defining how they manage information and handle external interactions, such as data fetching or user input (e.g., mouse clicks)~\cite{reacthooks}.
Among various built-in \hooks, we focus on two fundamental \hooks: \state and \effect.
These two \hooks are introduced early in the official \react tutorials~\cite{reactUseState, reactUseEffect} and are essential for understanding both the life cycle of \components~\cite{reactLifecycle} and their rendering process~\cite{reactRender}.
In fact, our analysis of \textit{Stack Overflow}\footnote{\url{https://stackoverflow.com/}} posts from 2020 to 2024 reveals that questions about these two \hooks account for 82\% of all \hooks-related posts, indicating that developers frequently struggle to use them (\autoref{fig:stackoverflow}).
The two \hooks are described as follows:

\begin{figure}[t]
    \centering
    \includegraphics[width=0.9\linewidth]{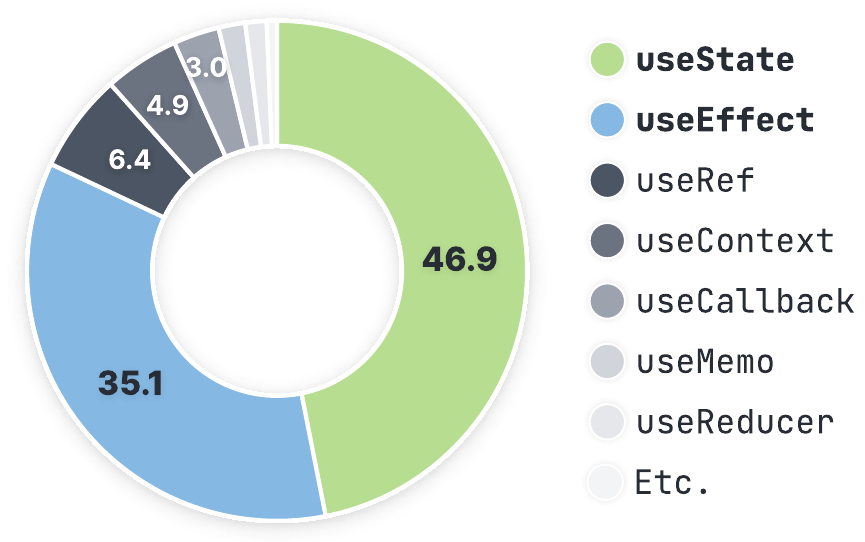}
    \caption{Ratio of posts about each built-in \hook on \textit{Stack Overflow} (2020–2024). Data are obtained by querying posts tagged with \textit{reactjs} that mention the name of each \hook using the \textit{Stack Exchange API} (\url{https://api.stackexchange.com/}). \useState and \useEffect denote the functions for the \state and \effect \hooks, respectively.}
    \label{fig:stackoverflow}
    \vspace{-0.2cm}
\end{figure}

\textbf{\state \hook}, which can be defined using a function named \useState, manages data within a \component.
The value stored in a \state \hook can be accessed within the corresponding \component and passed to its child \components through \props.
When this value is modified by external sources, such as user interactions, the \hook triggers a re‑render of the \component and its child \components to reflect the update.
In this way, the \state \hook enables developers to build dynamic and interactive applications that respond to data changes.

\textbf{\effect \hook}, provided by the \useEffect function, defines how a \component responds to changes in external sources.
This \hook allows developers to specify the operations to perform (logic) and the conditions that trigger them (dependencies), which typically depend on the values of \state \hooks or \props.
For example, when a value of \state \hook included in the dependencies of an \effect \hook changes, the associated logic is executed.
The \effect \hook thus enables \react applications to perform operations beyond re-rendering, such as synchronizing data with other \components or external sources.

We refer to \state \hook and \effect \hook simply as \state and \effect hereafter.

\subsection{Anti-Patterns in React}
\label{sec:antipatterns}
Although \hooks are powerful for developing interactive and dynamic applications, unmanaged or excessive use of them often leads to undesirable complexities and problems, known as anti-patterns~\cite{FERREIRAPropDrilling, FardJavaScriptCodeSmells}.
Various anti-patterns are identified in \react development~\cite{FardJavaScriptCodeSmells, NguyenCodeSmellsWeb, FERREIRAPropDrilling, qiu2024react}, but we mainly focus on three major anti-patterns related to \states and \effects~\cite{reactcontext, reacteffect}.
The following are descriptions of these three anti-patterns (\autoref{fig:antipatternexample}):

\begin{figure*}[ht]
    \centering
    \includegraphics[width=0.96\textwidth]{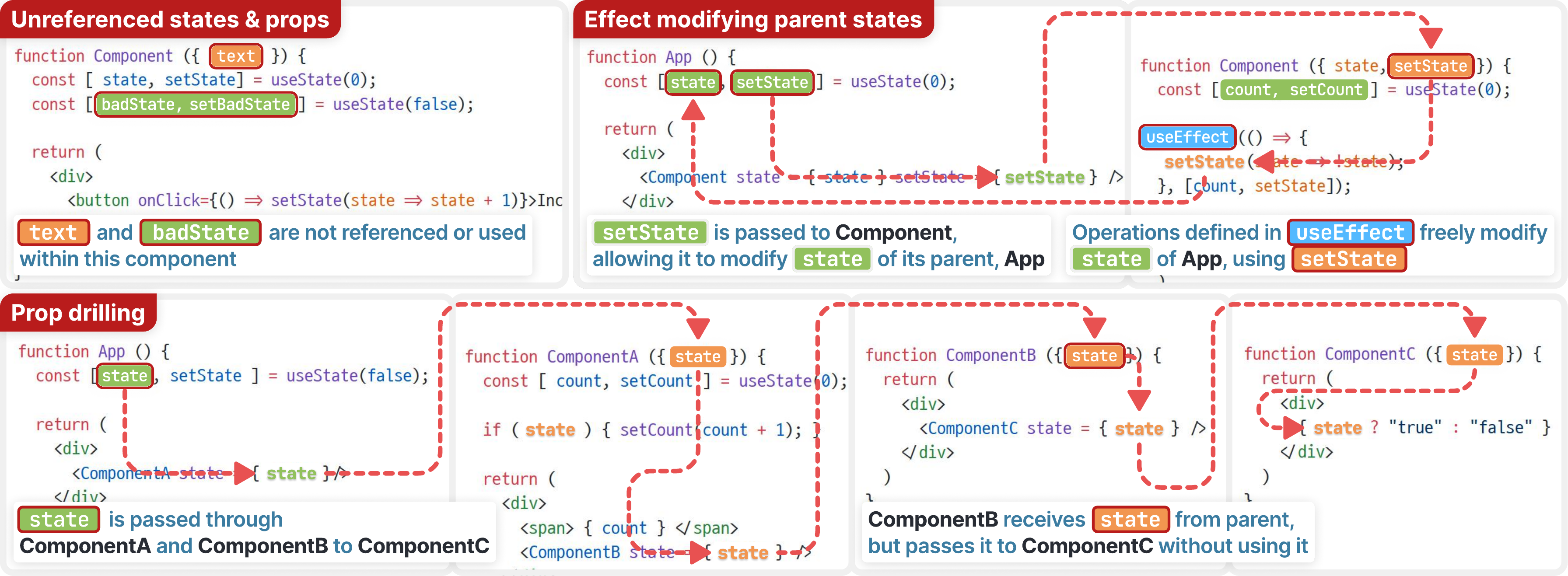}
    \caption{Examples of three anti-patterns. Each code snippet illustrates \textit{Unreferenced States and Props}, \textit{Effect modifying parent States}, and \textit{Prop drilling}, respectively. In each example, relevant \states, \effects and \props within \components are highlighted with red outlines, while red edges across \components indicate \state propagation paths.}
    \label{fig:antipatternexample}
    \vspace{-0.3cm}
\end{figure*}

\vspace{4pt}
\noindent
\textbf{\textit{Unreferenced States and Props.}} 
This anti-pattern denotes a code pattern in which \state values or \props are either unused within their defining \components or merely passed to another \component without being utilized.
This pattern can increase memory consumption and introduce unnecessary code, potentially leading to additional anti-patterns.

\vspace{4pt}
\noindent
\textbf{\textit{Prop drilling.}}
This pattern occurs when \state values are passed through multiple \components without being directly used within them.
It introduces unnecessary dependencies and data flows among \components, making the code more difficult to maintain and the flow of data harder to trace.

\vspace{4pt}
\noindent
\textbf{\textit{Effect modifying parent States.}}
This pattern refers to code in which the \state values of a parent \component are modified by the execution of an \effect defined in one of its child \components.
Such a pattern complicates the understanding of the application's behavior due to unpredictable \state values changes, often leading to unintended outcomes or bugs.

%% file: REVISION/3_related_works.tex
\section{Related Work}
\label{sec:related_works}

We discuss prior studies related to \hooklens.
We first review research on analyzing \react applications from various perspectives, followed by studies on software visualization that aim to analyze and understand code.

\subsection{React Application Analysis}
\label{sec:reactapp}

\react has become widely used for web development, attracting research attention.
For example, Sharma et al.~\cite{SharmaVsAngular} compare the performance of \react with other frameworks such as \textit{Angular}, while \textit{ReactAppScan}~\cite{GuoReactAppScan} introduces a tool for detecting security vulnerabilities. 
Despite these efforts, our understanding of addressing the complexity arising from the architecture of \components and \hooks remains limited, particularly in supporting maintenance tasks.
To the best of our knowledge, \textit{React-tRace}~\cite{lee2025reacttrace} is the only attempt to address this gap.
It analyzes the semantics of \react code, focusing on the rendering process to support UI debugging.
However, interpretability remains an issue, as users cannot readily locate the specific code where the problems originate.
Similarly, though studies analyze \react code to identify and mitigate anti-patterns~\cite{FERREIRAPropDrilling, FERREIRAMigrateClassToFunction, NUNESTScodesmell}, detecting anti-patterns across multiple files---particularly those involving \hooks---remains challenging.
Visual analytics approaches, such as \textit{React Bratus}~\cite{BoersmaReactBratus} and Tarner et al.~\cite{TarnerCodeQuality}, aim to address this limitation by visualizing hierarchical \component structures, but developers still struggle to understand the behavior of \hooks and the dynamic nature of \react applications.

\textit{Our contribution.}
We propose a visual analytics system that assists developers in understanding code structures and detecting anti-patterns in \react applications.
Unlike prior studies~\cite{BoersmaReactBratus, TarnerCodeQuality} that primarily focus on \component structures or execution outcomes, our approach centers on the interactions among \components and \hooks, similar in spirit to \textit{React-tRace}~\cite{lee2025reacttrace}.
However, rather than analyzing rendering semantics, we focus on uncovering anti-patterns that emerge within \react environments shaped by \hooks.
Through this approach, developers can gain a deeper understanding of how \hooks influence the behavior of \react applications and more effectively explore anti-patterns related to \hooks.

\subsection{Software Visualization}
\vspace{4pt}
\noindent
\textbf{The utility of software visualization.}
Visualization and visual analytics are widely used to analyze, debug, and maintain software~\cite{chotisarn2020systematic, LiuVAMaintain, kim21tvcg, kim21pvis}.
Prior studies demonstrate that software visualization helps developers understand concepts that bind to certain environments or programming languages.
For example, \textit{RustViz}~\cite{AlmeidaRustViz} visually illustrates difficult-to-understand concepts in \textit{Rust}, such as \textit{ownership} and \textit{borrowing}.
\textit{InterLink}~\cite{LinInterLink} reveals the relationship between code and text cells in computational notebooks (e.g., Jupyter).
Software visualization can also uncover hidden information that is hard to perceive in raw source code, such as module dependencies or data flows.
For instance, Gouveia et al.~\cite{GouveiaUsingHTML5} represent the execution counts for each code segment using visual diagrams to support runtime debugging.
\textit{GraphBuddy}~\cite{BorowskiGraphBuddy} and \textit{DeJEE}~\cite{ShatnawiAnalyzingProgram} visualize the dependencies between modules and libraries in \textit{Java} environments.
Similarly, \textit{Reactive Inspector}~\cite{SalvaneschiDebug} and \textit{RxFiddle}~\cite{BankenDebugDataReactive} provide visual debugging tools that visualize signal flows and dependencies in reactive programming environments.

\vspace{4pt}
\noindent
\textbf{The utility of node-link diagrams in software visualization.}
While prior studies employ diverse visual idioms for software visualization, node-link diagrams stand out as a particularly effective and widely adopted approach.
Many of the aforementioned systems either directly employ node-link diagrams~\cite{BorowskiGraphBuddy, ShatnawiAnalyzingProgram, SalvaneschiDebug,BankenDebugDataReactive} or incorporate them into their visual designs~\cite{AlmeidaRustViz, LinInterLink}.
This is because node-link diagrams intuitively reveal relationships such as dependencies and data or control flows among program elements (e.g., functions, classes, and modules)~\cite{nodelinkdiagram}.
For instance, \textit{E-Quality}~\cite{TarnerCodeQuality} visualizes code quality metrics on node-link diagrams by following relationships among classes to support refactoring and assess maintainability.
In contrast, \textit{REACHER}~\cite{viscall} and Ishio et al.~\cite{visdata} employ node-link diagrams to assist program comprehension by revealing control and data flows within the source code.
Beyond traditional programming environments, \textit{Rapsai}~\cite{DuRapsai} and \textit{Misty}~\cite{LuMisty} extend this approach to higher-level development tasks such as constructing machine learning pipelines and prototyping mobile UIs.
More recently, node-link diagrams are employed to support LLM-based code generation, visualizing user prompts instead of source code to help users interpret and guide the code generation process~\cite{EarleDreamGarden, NeuroSync, sparkGenUI}.

\textit{Our contribution.}
Building on the proven utility of visualization in software analysis and maintenance, we employ software visualization---particularly node-link diagrams---to support code comprehension and anti-pattern detection based on \components and \hooks.
As prior works demonstrate the effectiveness of node-link diagrams across diverse software analysis, development and maintenance tasks, we extend conventional node-link diagrams into a nested and interactive structure.
Our approach enables developers to more effectively explore the hierarchical structure of \components and uncover hidden dependencies introduced by \hooks within these hierarchies.

%% file: REVISION/4_tasks_and_design_requirements.tex
\section{Tasks and Design Requirements}
\label{sec:challenges}

We introduce the tasks and design requirements of \hooklens.

\subsection{Tasks}
\label{sec:tasks}
We identify two core tasks through a semi-structured preliminary interview.
We recruit three \react developers (three males, aged 28, 30, and 34) with over two years of experience and intermediate proficiency.
\revised{All participants have experience working on at least two projects, either individually or in small teams of up to four members.}
We ask participants about their use of \hooks---particularly \state and \effect---in maintaining and refactoring code.
Each interview lasts up to 35 minutes.

\vspace{4pt}
\noindent
\textbf{(T1) Managing States and their propagation among Components.} 
\states within each \component are closely tied to its execution logic and rendering outcomes, and child \components may also depend on certain \states from their parent \components for their own execution.
While \states serve as essential and powerful tools for creating dynamic and interactive \components, developers often fail to manage them properly or tend to overuse them.
In particular, developers often propagate \state values and their setter functions redundantly across multiple child \components for convenience (i.e., \textit{prop drilling})
Such unmanaged and widely propagated \states make it difficult to trace their origins, and as the number of \components increases, even minor code changes may require reviewing or modifying a large number of \components, thereby reducing maintainability.
Therefore, developers need to understand where \states are defined, how they are propagated, and which \components are affected by their changes.

\vspace{4pt}
\noindent
\textbf{(T2) Tracking the chains of Effects.} 
As the logic defined in \effects can execute independently from the rendering process of \react, developers need to track and understand how these \effects are executed.
However, when a \component handles too many functionalities or becomes overly complex, excessive use of \effects may occur, forming long chains of \effects.
In some cases, these \effects even propagate upward through the \component hierarchy, affecting parent \components (i.e., \textit{Effect modifying parent states}).
Since these \effects depend on runtime conditions and are triggered by external sources across multiple \components, predicting their outcomes is challenging.
Such unpredictability can introduce unintended behaviors that are difficult to trace and fix, and can further degrade application performance.
Therefore, developers should carefully track and understand the behaviors and outcomes defined by these \effects.

\subsection{Design Requirements}
\label{sec:requirements}

We formulate design requirements to support developers in effectively performing the core tasks identified in \autoref{sec:tasks}.

\vspace{4pt}
\noindent
\textbf{(DR1) Focus on Props, States, and Effects within Components.} 
Although diverse approaches exist for analyzing \react applications~\cite{BoersmaReactBratus, TarnerCodeQuality}, developers primarily reason about \component behavior through \props, \states, and \effects (T1-–T2).
This is because \states and \effects play crucial roles in determining application behavior (e.g., rendering process and \component life cycle), while \props serve as essential links to trace data and control flow across \components.
Therefore, \hooklens should focus on these three elements to help developers understand data flow, debug logic, and identify dependencies.

\vspace{4pt}
\noindent
\textbf{(DR2) Deliver the hierarchical structure of Components.} 
\react applications operate and render based on the hierarchical structures of \components, where \state values propagate along these hierarchies through \props.
To understand such operations and propagation (T1), developers need to examine the hierarchical relationships among \components.
However, these hierarchies are often deeply nested and difficult to infer from source code alone.
Therefore, \hooklens visually represents these hierarchical structures to help developers comprehend how data and control flow through the \component hierarchy.

\vspace{4pt}
\noindent
\textbf{(DR3) Explain the dynamic interactions among Props, States, and Effects.} 
Beyond identifying these core elements (DR1), understanding how they dynamically interact is key to reasoning about \react application behavior. 
In \react applications, data and control flows emerge from the interactions among \props, \states, and \effects (\autoref{sec:hooks}). 
For example, updates to \states of the parent \component may propagate as \props to child \components (T1), triggering \effects that perform additional operations (T2). 
Therefore, \hooklens should visualize these dynamic dependencies to help developers understand how these elements influence the overall behavior of applications.

\vspace{4pt}
\noindent
\textbf{(DR4) Reveal potential anti-patterns.} 
While anti-pattern detection is crucial for refactoring~\cite{maintenence, fowler2018refactoring}, it remains challenging in \react code, especially as applications scale in size and complexity.
In particular, anti-patterns related to \hooks often span multiple \components and files (e.g., \textit{prop drilling} and \textit{Effect modifying parent states}), making them difficult to detect and trace.
These complexities hinder issue resolution and reduce code maintainability.
Therefore, \hooklens should highlight problematic patterns in the use of \hooks to help developers identify and address potential issues and refactoring opportunities.

%% file: REVISION/5_hooklens.tex
\section{Iterative Design Process}

We adopt an iterative design process to develop \hooklens. 

\vspace{4pt}
\noindent
\textbf{Designing the initial prototype.} 
We first build an initial prototype based on the design requirements (\autoref{sec:requirements}) to explore potential improvements in visual representations and interactions.
Our prototype visualizes the hierarchical structure of \components (DR2) and the relationships among \states and \effects (DR1, DR3) through a node-link diagram. 
Moreover, the prototype can automatically detect and visually highlight \textit{unreferenced states and props} anti-patterns (DR4).

\vspace{4pt}
\noindent
\textbf{Receiving feedback from experienced \react developers.}
We conduct feedback sessions with eight experienced \react developers (eight males, aged 22--31 [27$\pm$5]), each with more than two years of experience.
During each session, we explain the goal of \hooklens to the participants and demonstrate its functionality using a sample project that contains several anti-patterns, such as \textit{unreferenced states and props} and \textit{prop drilling}.
We then present a simple scenario in which participants trace \states within a specific \component and identify the dependencies of an \effect.
Throughout the session, participants are encouraged to freely share their opinions and suggestions regarding the prototype.

\vspace{4pt}
\noindent
\textbf{Designing the final prototype.}
Based on the feedback, we refine the final prototype of \hooklens with two major improvements.

\vspace{4pt}
\noindent
\textit{(I1) Highlight areas of interest.} 
While most participants agree that the prototype effectively reveals the structural relationships among \components, \states, and \effects, they report difficulties in keeping track of which elements they are focusing on during exploration.  
This issue becomes particularly critical in larger projects, where the large number of nodes and edges makes navigation more challenging.
To address this, we enable users to interactively select and visually emphasize areas of interest. 

\vspace{4pt}
\noindent
\textit{(I2) Integrate a code viewer for detailed analysis.} 
Participants emphasize the need for an integrated code viewer to support detailed analysis of the application.
Although they can identify \components or \hooks where issues exist or validation is required, the prototype alone makes it difficult to review exact execution timings or data changes.
In particular, some participants suggest adding a feature that directly displays the code related to the focused \components or \hooks.  
We therefore integrate a code viewer that highlights the code corresponding to the elements currently focused in the visualization.

\section{HookLens: An Interactive Visual Analytics System}
\label{sec:hooklens}

We present \hooklens, an interactive visual analytics system that helps developers understand the overall structure of \react applications through \components and \hooks, and identify related anti-patterns.
The system extracts key elements such as \components, \props, and \hooks from a given \react project and visualizes their relationships through an interactive node-link diagram.

\begin{figure}[t]
    \centering
    \includegraphics[width=\linewidth]{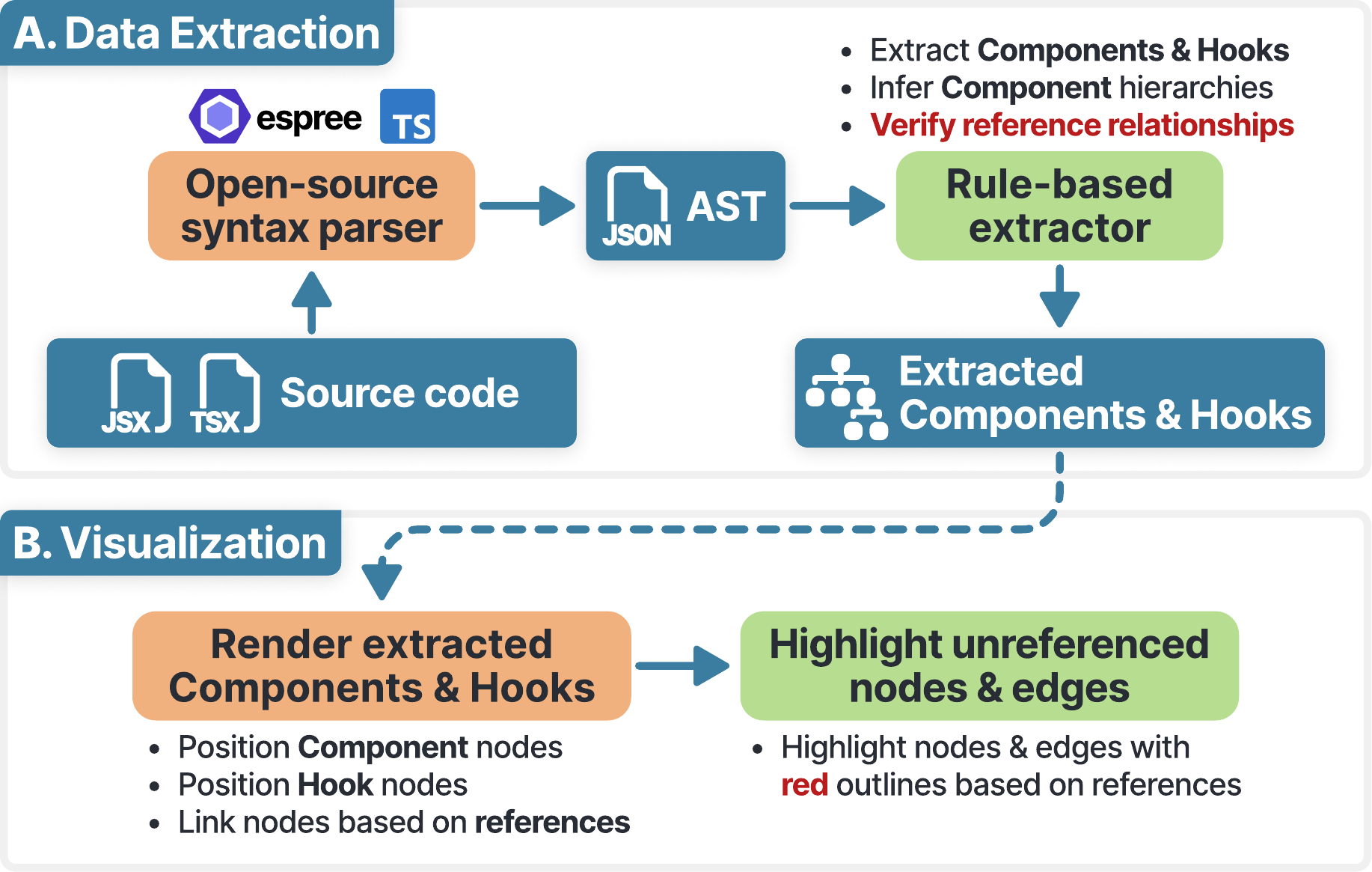}
    \caption{
    Data extraction and visualization pipeline of \hooklens. \hooklens parses source files into abstract syntax trees (ASTs) using \textit{Espree} and the \textit{TypeScript} compiler API, then extracts \components, \hooks, and their reference relationships with predefined rules. It visualizes the extracted data as a node-link diagram, highlighting invalid references in red to facilitate anti-pattern detection.}
    \label{fig:dataprocessing}
    \vspace{-0.1cm}
\end{figure}

\subsection{Data Extraction Pipeline}
Given a \react project, \hooklens extracts \components, \props, and \hooks (i.e., \states and \effects) and builds structured data based on the relationships among these elements (DR1; \autoref{fig:dataprocessing}A).
To ensure fast and consistent extraction of these elements, we develop a rule-based extractor module.
We first leverage (1) the compiler API of \textit{TypeScript}~\cite{tscompiler} and (2) the open-source \textit{JavaScript} parser, \textit{Espree}~\cite{espree}, to convert the source code into abstract syntax trees (ASTs).
We adopt ASTs because they are widely used for static code analysis~\cite{AST1, TurcotteDrAsync} and facilitate systematic exploration of source code.
Building on these ASTs, we implement extraction logic that detects \components, \props, \states, and \effects, and constructs their relationships.
This logic follows predefined rules to identify target elements and trace their dependencies.
\revised{Furthermore, the extractor verifies reference relationships among the extracted elements to identify unreferenced or unused \components, \props, and \states, as well as \effects that introduce backward dependencies to their parent \components.}
As a result, the extractor module builds a graph-structured dataset that represents the extracted elements and their relationships.

\subsection{Visual Analytics Design}
We describe how \hooklens visualizes code structures and anti-patterns related to \hooks \revised{(\autoref{fig:dataprocessing}B)}, and how it supports interactive exploration (\autoref{fig:teaser}).

\vspace{4pt}
\noindent
\textbf{Visualization for Components and Hooks.}
\hooklens visualizes the extracted \components, \props, and \hooks, along with their relationships, using an interactive node-link diagram.
Since node-link diagrams align with how developers conceptualize function calls and data flows~\cite{nodelinkdiagram}, \hooklens adopts this approach to facilitate users' understanding of \state value propagation (T1) and chains of \effects (T2).
Nodes in the diagram represent extracted elements---\components, \props, \states, and \effects (DR1)---while edges indicate relationships such as dependencies of \effects and data flows between \states and \props (DR3).
In particular, \state, \prop, and \effect nodes are placed within their corresponding \component node to indicate that these elements are defined by the \component.
To clearly distinguish different types of nodes and edges, we use distinct hues with consistent saturation levels, thereby avoiding unintended implications of ordinal attributes~\cite{TsengCategorical, manckinlayHueSaturation}.
Furthermore, \hooklens highlights potential anti-patterns (\autoref{fig:antipattern}) using animated red outlines and edges (DR4; \autoref{fig:teaser}B).
\revised{Based on the reference relationships verified during the extraction process (\autoref{fig:dataprocessing}), the system emphasizes unreferenced or unused elements as well as \effects that introduce backward dependencies to their parent \components.
These visual cues make potential anti-patterns visually salient, as all target anti-patterns in this study arise from reference relationships and their associated data or control flows.}
We adopt red as it is a widely recognized visual cue for warnings in development environments.

\vspace{4pt}
\noindent
\textbf{Interactive exploration.}
To help users explore relationships among \hooks without cognitive overload, we design \hooklens following Shneiderman's visual information seeking mantra~\cite{ShneidermanTheeyeshaveit}.
The system first presents an overview of \components and their relationships, providing a high-level understanding of the project.
\component nodes are arranged horizontally by hierarchy, enabling developers to view deeply nested structures within a single screen (DR2).
Users can then reveal \states, \effects, \props, and their relationships on demand.
Clicking on a \component node expands it to show its internal \states, \effects, and \props (\autoref{fig:teaser}A and B).
When related \components are also expanded, the system visualizes edges representing the propagation of \state values and \props (DR3).
Clicking a \state, \effect, or \prop node highlights its related data and control flows while dimming others (I1; \autoref{fig:teaser}C).
Users can further navigate the diagram using pan and zoom interactions.
For deeper inspection, \hooklens integrates a code viewer that automatically locates and highlights the source code corresponding to the selected node (I2; \autoref{fig:teaser}D).

%% file: REVISION/6_evaluation.tex
\section{User Study}
\label{sec:user_study}

We conduct a user study to evaluate the effectiveness and usability of \hooklens in performing the target tasks (\autoref{sec:tasks}), comparing it with a popular code editor (\textit{VS Code}).

\subsection{Objectives and Design}

We aim to evaluate how effectively \hooklens supports developers in performing the target tasks (\autoref{sec:tasks}) and how usable the system is for understanding and maintaining \react applications.
To this end, we recruit \react developers and measure their performance in terms of task accuracy.
We further assess the usability of the system through the \textit{System Usability Scale} (SUS) and post-study interviews.

\vspace{4pt}
\noindent 
\textbf{Participants.}
We recruit 12 \react developers (eight males and four females, aged 21--30 [25$\pm$9]) through snowball sampling.
In particular, we recruit participants with diverse levels of expertise \revised{in \react development} to examine whether \hooklens effectively supports both novice and intermediate developers.
Six participants (P1--P6) have less than two years of experience, while the remaining six (P7--P12) have more than two years of experience, consistent with the conditions of the preliminary interviews and feedback sessions. 
All participants are compensated with the equivalent of USD 10.

\vspace{4pt}
\noindent
\textbf{Tasks.}
We ask participants to detect anti-patterns described in \autoref{sec:antipatterns} within a \react project.
Since detecting anti-patterns is a common task in code maintenance and refactoring~\cite{maintenence, fowler2018refactoring}, we adopt it as the basis of our study.
Participants are asked to identify as many anti-patterns related to the use of \states and \effects as possible within a 10-minute time limit.
Before each session, we inform participants about the three target anti-patterns described in \autoref{sec:antipatterns} and instruct them to detect only these patterns.
Participants verbally report the presence of each anti-pattern and specify the associated \components, \props, \states, and \effects.
We measure precision and recall in detecting anti-patterns, using a ground truth established through careful code examination.
\revised{To ensure a fair comparison and focus on the core capabilities of \hooklens, we consider only reports involving \states and \effects when computing these metrics.
To ensure objectivity, the ground truth is constructed based on the formal definitions and code-level criteria of each anti-pattern (\autoref{sec:antipatterns}) and validated through independent review and consensus between two authors.
Reports related to other variables (e.g., variables from \texttt{useRef} or general \textit{JavaScript} variables) are excluded from the evaluation, as participants are instructed in advance to focus on the target patterns.}

\vspace{4pt}
\noindent 
\textbf{Baseline.}
\revised{To evaluate the cognitive advantages of visual analytics over text-based exploration, we select \textit{VS Code} as the baseline.
In addition to being one of the most widely used development tools~\cite{soPopularTech}, all preliminary interviewees and 11 out of 12 study participants report using \textit{VS Code} together with extensions such as \textit{eslint-plugin-react-hooks}\footnote{\url{https://react.dev/reference/eslint-plugin-react-hooks}}.
However, \textit{eslint-plugin-react-hooks} primarily enforces basic implementation rules—--such as ensuring consistent \hooks call order--—rather than detecting complex anti-patterns that span multiple components.
As a result, identifying such anti-patterns in text-based environments often requires extensive manual tracing, which imposes a significant cognitive burden.
Therefore, to compare \hooklens against this manual process, we provide a baseline environment with only the default \textit{IntelliSense} features enabled, allowing participants to use standard navigation functions (e.g., \textit{Go to Definition} and \textit{Find All References}) to manually trace dependencies.}

\vspace{4pt}
\noindent 
\textbf{Sample projects.}
We select two open-source projects (\textit{Confides}~\cite{ConfidesURL} and \textit{paper\_vis}~\cite{papervis}) to simulate real-world tasks.
\revised{Many popular open-source \react projects are developed by large teams and undergo extensive maintenance, during which such anti-patterns are often controlled or resolved, making them less suitable for our study tasks within a limited time budget.
Therefore, we focus on projects developed by individuals or small teams, which are academic prototypes.
Such projects are typically developed as proof-of-concept systems and thus tend to contain a sufficient number of representative anti-patterns, making them well suited for controlled evaluation of anti-pattern detection.}
To mitigate learning effects that could bias tool performance, we employ two separate projects with comparable task complexity, each containing a similar number of files, \components, and anti-patterns (\autoref{tab:project_stats}).

\begin{table}[t]
    \centering
    \caption{Comparison of two open-source projects used in the study. We select the projects that are similar in terms of their size and the number of existing anti-patterns.}
    \begin{tabular}{lcc}
        \toprule
        \textbf{Metric} & \textit{Confides} & \textit{paper\_vis} \\
        \midrule
            \multicolumn{3}{l}{\textbf{Size}} \\
            \quad Number of JSX files & 29 & 30 \\
            \quad Number of components & 25 & 33 \\
            \quad Total lines of code (JSX) & 2707 & 3937 \\
        \midrule
            \multicolumn{3}{l}{\textbf{Anti-patterns}} \\
            \quad Unreferenced states \& props & 41 & 32 \\
            \quad Prop drilling & 11 & 11 \\
            \quad Effect modifying parent states & 2 & 2 \\
        \bottomrule
    \end{tabular}
    \label{tab:project_stats}
\end{table}

\begin{figure}[t]
    \centering
    \includegraphics[width=\linewidth]{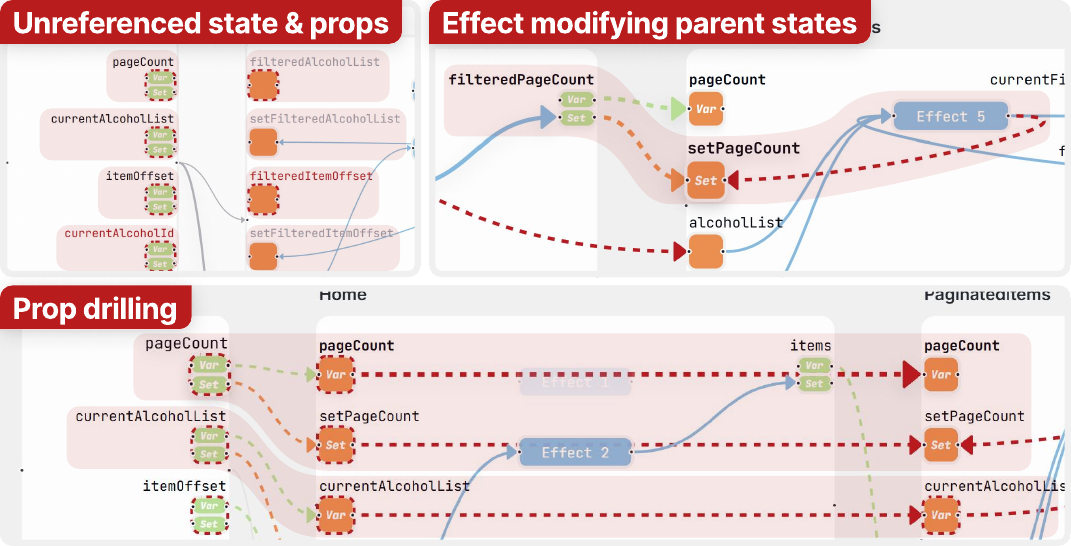}
    \caption{Types of anti-patterns revealed by \hooklens. Anti-patterns are highlighted with translucent red backgrounds: \textit{Unreferenced states and props}, shown as red outlined \state and \prop nodes; \textit{prop drilling}, represented by red edges between \prop nodes; and \textit{effect modifying parent states}, represented by red edges originating from \effect nodes.}
    \vspace{-0.1cm}
    \label{fig:antipattern}
\end{figure}

\vspace{4pt}
\noindent 
\textbf{Procedure.}
After participants consent to their participation, we first explain \hooklens and its key features (e.g., visualization and available interactions).
We then provide participants approximately 20 minutes to explore and practice with the system, where they are allowed to freely ask questions.
After the practice session, we spend about 5 minutes introducing the study tasks and the three target anti-patterns (\autoref{sec:antipatterns}), presenting examples of each in both code (\autoref{fig:antipatternexample}) and visualization (\autoref{fig:antipattern}) within \hooklens.
Subsequently, participants perform the task in two conditions: using \textit{VS Code} and \hooklens, with each condition lasting 10 minutes (20 minutes in total).
To control for ordering effects, half of the participants begin with \hooklens, while the others start with \textit{VS Code}.
We also assign different projects for each condition to control learning effects.
After completing both sessions, participants fill out the SUS questionnaire and participate in a post-study interview for approximately 15 minutes to provide qualitative feedback on system usability and overall satisfaction.
In total, the entire study session takes approximately one hour.

\definecolor{appleredlight}{RGB}{255, 105, 97}
\newcommand{\accolor}[1]{\cellcolor{appleredlight!#1}}

\begin{table*}
    \centering
    \caption{Performance comparison between \hooklens and \textit{VS Code} in detecting different anti-patterns. Values are reported as mean $\pm$ standard deviation. All results are statistically significant ($p \ll 0.01$). We color the table cells in red with an opacity scale where lower opacity represents lower accuracy (linear gradient between 0 and 1).}
    \begin{tabular}{l cccccc}
        \toprule
        \multirow{2}{*}{Anti-pattern} & \multicolumn{3}{c}{\textbf{\hooklens}} & \multicolumn{3}{c}{\textbf{\textit{VS Code}}} \\
        \cmidrule(lr){2-4} \cmidrule(lr){5-7}
         & Precision & Recall & F1 & Precision & Recall & F1 \\
        \midrule
        Unreferenced states \& props 
        & \accolor{96.8}.968 $\pm$ .098 & \accolor{50.9}.509 $\pm$ .311 & \accolor{61.4}.614 $\pm$ .272 
        & \accolor{66.9}.669 $\pm$ .322 & \accolor{14.7}.147 $\pm$ .111 & \accolor{21.9}.219 $\pm$ .153 \\
        
        Prop drilling 
        & \accolor{93.8}.938 $\pm$ .121 & \accolor{56.8}.568 $\pm$ .255 & \accolor{66.9}.669 $\pm$ .213 
        & \accolor{49.2}.492 $\pm$ .457 & \accolor{19.7}.197 $\pm$ .243 & \accolor{26.4}.264 $\pm$ .294 \\
        
        Effect modifying parent states
        & \accolor{88.9}.889 $\pm$ .283 & \accolor{79.2}.792 $\pm$ .320 & \accolor{81.7}.817 $\pm$ .284 
        & \accolor{18.1}.181 $\pm$ .369 & \accolor{16.7}.167 $\pm$ .312 & \accolor{16.0}.160 $\pm$ .316 \\
        \bottomrule
    \end{tabular}
    \vspace{-0.05cm}
    \label{tab:antipattern_results}
\end{table*}

\begin{figure*}[t]
    \centering
    \includegraphics[width=\textwidth]{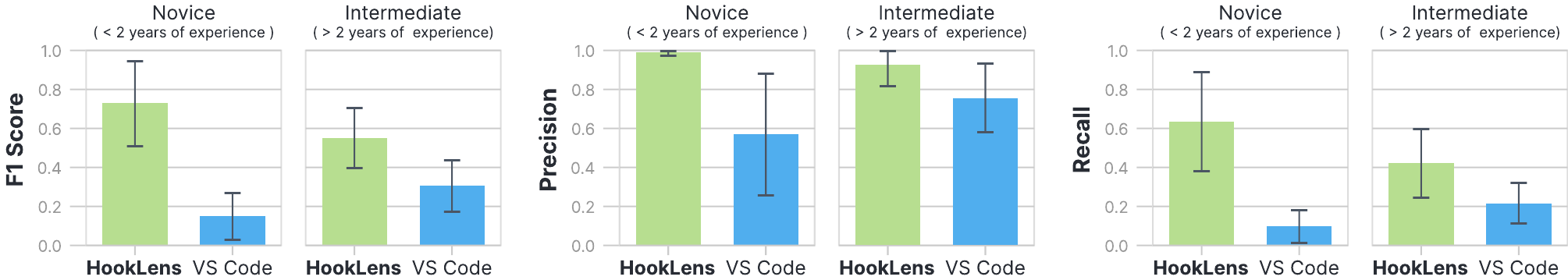}
    \caption{Accuracy comparison between \hooklens and \textit{VS Code} in detecting anti-patterns, separated by proficiency. Novice participants have less than two years of experience, while intermediate participants have more than two years. Using \hooklens resulted in statistically significant improvements in F1-score and recall for \revised{both novice ($p \ll 0.01$) and intermediate ($p \ll 0.05$)} groups, though not in precision.}
    \label{fig:groupresult}
    \vspace{-0.2cm}
\end{figure*}

\subsection{Quantitative Results}
\label{sec:quantAccuracy}

Our study results confirm the effectiveness of \hooklens in supporting participants in detecting anti-patterns.
They further imply the superiority of \hooklens in user experience compared to a conventional code editor.

\vspace{4pt}
\noindent
\textbf{Accuracy in detecting anti-patterns.} 
We calculate F1-scores for anti-pattern detection based on participants' responses in tasks performed with both \textit{VS Code} and \hooklens.
A Mann–Whitney U test on these scores shows that participants achieve significantly higher task accuracy with \hooklens than with \textit{VS Code} (\autoref{tab:antipattern_results}).
Further statistical analysis by expertise level suggests that \hooklens effectively supports both novice and intermediate developers in identifying anti-patterns, although differences between the two groups are also observed (\autoref{fig:groupresult}).
We further discuss these differences between novice and intermediate developers in the qualitative findings section (\autoref{sec:qualitative}).

\vspace{4pt}
\noindent
\textbf{Usability of \hooklens.} 
The SUS results indicate that \hooklens shows high usability for the given tasks, with an average score of 76.7. 
This score exceeds the standard benchmark of 68, suggesting that participants generally found the system easy to learn and use~\cite{SUSStandard}. 
These findings are also consistent with the qualitative findings obtained from the post-study interviews (\autoref{sec:qualitative}).

\subsection{Qualitative Findings}
\label{sec:qualitative}

We present the qualitative findings derived from task observations and post-study interviews.
These findings provide plausible explanations for the quantitative outcomes and further highlight both the strengths and potential areas for improvement of \hooklens.

\vspace{4pt}
\noindent 
\textbf{Advantages of visual analytics in understanding React applications.} 
Most participants agree that the visualization provided by \hooklens is highly effective for understanding the overall structure and the relationships among \components and \hooks across multiple files, which are often difficult to grasp with conventional code editors.
For example, several participants (e.g., P2, P9, P10) mention that \hooklens helps them quickly gain an overview of unfamiliar projects, as its visualization supports comprehension without requiring line-by-line code inspection.
P2 adds that the spatial positioning of \component nodes intuitively conveys structural hierarchy, making it easier to interpret the layout.
P9 notes that this advantage could substantially support refactoring legacy code within their organization.
Several participants (e.g., P3, P6, P8, P12) also highlight that \hooklens facilitates identifying specific files and \components relevant to their tasks, even when multiple \components are defined in a single file or when directory structures are complex.
Participants further report that \hooklens is effective in tracing \state propagation and \effect dependencies across \components.
P7 notes that it helps investigate how \components and \states reference each other and which \components are designed for shared use across \components.
Similarly, P3 and P4 mention its usefulness for managing \states, such as locating where they are defined and which \props deliver their values, while P8 finds that tracing \effects is straightforward by simply following the directed edges.
Additionally, P9 and P12 emphasize that \hooklens makes it much easier to trace relevant \components, \states, and \effects compared with conventional editors, which require developers to manually search across multiple files.

\vspace{4pt}
\noindent 
\textbf{Usability in detecting anti-pattern.}
Both novice (e.g., P1, P3, P6) and intermediate (e.g., P8, P11, P12) participants find that \hooklens is useful for identifying anti-patterns.
They note that the red highlights in \hooklens effectively draw their attention, enabling them to quickly recognize the relevant \components.
In particular, P6 and P11 emphasize that analyzing the relationships among \components and \hooks should take precedence over detailed source code analysis when detecting such anti-patterns, and that \hooklens is therefore especially helpful by filtering out unrelated \components.
Meanwhile, P3 notes that when he encounters \textit{prop drilling}, \hooklens facilitates exploration by visualizing the detailed references of \props and helping prioritize which ones may pose potential risks.

\vspace{4pt}
\noindent 
\textbf{Differences in use between novice and intermediate participants.}
As shown in \autoref{fig:groupresult}, novice participants perform better with \hooklens than intermediate participants, \revised{and the observed p-values indicate stronger statistical significance for the novice participants.}
During the study tasks with \hooklens, most novice participants rely heavily on the visualizations provided by \hooklens.
\revised{For example, P1 relies entirely on visual exploration and the visual cues provided by \hooklens to report answers, without directly inspecting the source code.
Similarly, most other novice participants (e.g., P2, P4--P6) primarily rely on visual exploration to complete the tasks rather than inspecting the source code, which leads to faster anti-pattern detection.
Exceptionally, P3 tends to inspect the source code more frequently than other novice participants, but consistently achieves low accuracy in both the baseline and \hooklens conditions.
Meanwhile, intermediate participants tend to use the visualization as a reference while repeatedly reviewing the source code throughout the tasks.
They generally achieve higher accuracy than novice participants in the baseline condition, suggesting that they are more familiar with navigating source code.
This greater familiarity is further reflected in their review of the source code even when performing tasks with \hooklens, which is associated with longer anti-pattern detection times compared to novice participants.
In fact, we observe that P8 and P9 keep the source-code view open throughout most of the task execution.
}

Notably, most intermediate participants (e.g., P8, P10--P12) note that, unlike \textit{VS Code}, \hooklens provides limited features for direct code inspection and detailed behavior verification, making it challenging to fully grasp application logic.
By contrast, novice participants (e.g., P1--P3) report that they can understand application behavior more easily, likely because they usually work on projects where the interactions among \components, \states, and \effects are relatively simple.

\vspace{4pt}
\noindent 
\textbf{Limitations in supporting real-world development workflows.}
Although \hooklens facilitates users in performing core tasks, participants also point out several limitations.
First, some participants note that although \hooklens is effective for examining code structures and identifying anti-patterns, it provides limited support for analyzing runtime behavior.
P8 emphasizes that detailed analysis becomes feasible only after gaining sufficient familiarity with the system, and P11 adds that \hooklens restricts source code exploration compared to \textit{VS Code}.
Several participants also suggest that \hooklens should include additional features to better support real-world development.
For example, P9 and P10 express concern that \hooklens does not support external state management systems such as \textit{Redux}, which are widely used in practice and can help mitigate \textit{prop drilling}.
Finally, P12 points out that tracing relationships in regions with many intersecting edges can be overwhelming, highlighting the need to improve visual scalability to better support larger projects.

%% file: REVISION/7_evalute_LLM_assistant.tex
\section{Comparative Evaluation with LLM-based Coding Assistants}
\label{sec:llmcompare}

To further assess the effectiveness of \hooklens, we compare its performance against state-of-the-art LLM-based coding assistants. 
Specifically, we evaluate how well these assistants identify anti-patterns related to \hooks and whether their detection performance is comparable to that of human-in-the-loop visual analytics.

\subsection{Objectives and Design}
We evaluate whether \hooklens outperforms state-of-the-art LLM-based coding assistants in detecting anti-patterns.
To this end, we select representative assistants and measure their task accuracy using the same procedure as our user study, comparing the results with those of the participants.

\vspace{4pt}
\noindent
\textbf{LLM-based coding assistants.}
We select three state-of-the-art LLM-based coding assistant tools to evaluate their capabilities in detecting anti-patterns.
To ensure fairness and consistency, we focus on cloud-based commercial tools that demonstrate high performance on the de facto benchmark, \textit{SWE-bench}\footnote{\url{https://www.swebench.com/}}~\cite{jimenez2024swebench}, which evaluates models on code understanding and software engineering problem solving.
The selected assistants are as follows:

\begin{itemize}[leftmargin=*, noitemsep, topsep=2pt]
    \item \textit{\textbf{Claude Code}}~\cite{claudecode} provides the \textit{claude-sonnet-4} (model: claude-sonnet-4-20250514, released May 2025) and \textit{claude-opus-4.1} (model: claude-opus-4-1-20250805, released Aug 2025) models, which currently achieve the highest scores on \textit{SWE-bench}.
    \item \textit{\textbf{Codex CLI}}~\cite{codexcli} offers one of the latest high-performing models, \textit{GPT-5} (released Aug 2025).
    \item \textit{\textbf{Gemini CLI}}~\cite{geminicli} provides the \textit{gemini-2.5-pro} (released Aug 2025) model, which performs slightly below the other two tools but still ranks highly on \textit{SWE-bench}.
\end{itemize}

\vspace{4pt}
\noindent
\textbf{Procedure.}
We conduct the same tasks as in the user study (\autoref{sec:user_study}) for each coding assistant and compare the results with those of human participants using \hooklens.
For each assistant, we perform six independent trials on the two projects used in the user study and count the number of detected anti-patterns using the same ground truth.
To ensure equivalent conditions, we grant each assistant full access to all project files, reset its state before each trial, and provide an identical prompt \revised{(appendix A)} that includes a code-level one-shot example of each anti-pattern, replicating the conditions given to user study participants.

\begin{figure*}[ht]
    \centering
    \includegraphics[width=\textwidth]{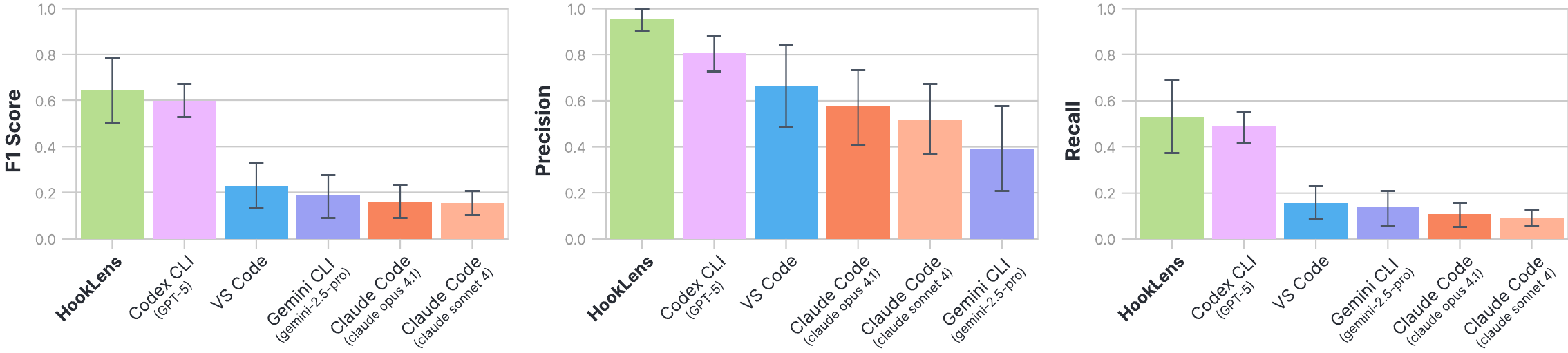}
    \caption{Accuracy comparison between human developers and LLM-based coding assistants in detecting anti-patterns. \hooklens shows substantially higher mean accuracy than other tools, which is statistically significant ($p \ll 0.01$) except in the case of \textit{GPT-5}. In terms of precision, \hooklens also outperforms \textit{GPT-5} ($p \ll 0.01$).}
    \label{fig:llmresult}
    \vspace{-0.3cm}
\end{figure*}

\begin{figure}[ht]
    \centering
    \includegraphics[width=\linewidth]{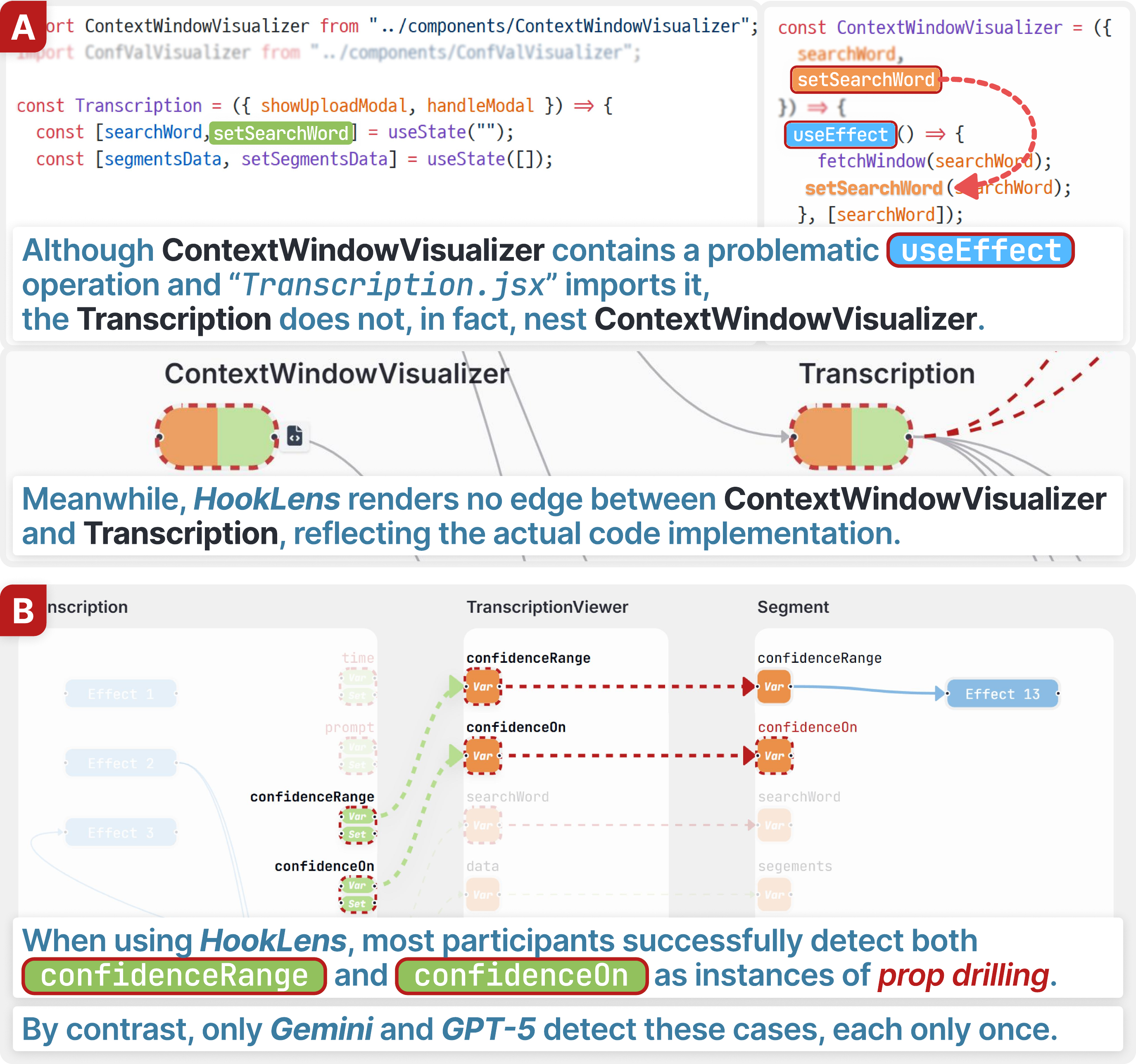}
    \caption{Falsely reported result within \textit{Confides} project by LLM-based coding assistants. LLMs detect the code snippet as \textit{effect modifying parent states}, while it does not produce anti-patterns because \texttt{Transcription} \component does note nest \texttt{ContextWindowVisualizer} \component in fact (A). \revised{Meanwhile, LLMs fail to detect \textit{prop drilling} patterns although it is detected by most participants with \hooklens in the user study (B).}}
    \label{fig:llmfailure}
    \vspace{-0.2cm}
\end{figure}

\subsection{Results}
\label{sec:llmresult}
Compared to participants using \hooklens, most LLM-based assistants struggle to detect anti-patterns (\autoref{fig:llmresult}).
Except for \textit{Codex CLI} with \textit{GPT-5}, all tools perform worse than even the \textit{VS Code} baseline. 
Although \textit{GPT-5} achieves relatively better performance than other LLMs, its precision remains statistically lower than that of \hooklens. 
These results are consistent with prior studies showing that LLMs often face difficulties in analyzing code with extensive context~\cite{WuIsmell, TessaSmell, DetectingCodesmellsGPT}.
In particular, the anti-patterns examined in this study---spanning multiple files and requiring contextual reasoning---further expose this limitation.
For example, we observe that all four LLMs frequently produce false positives by incorrectly identifying imported relationships that are never actually referenced (\autoref{fig:llmfailure}A).
\revised{We also observe that LLMs repeatedly fail to detect certain \textit{prop drilling} patterns that are identified by most participants using \hooklens (\autoref{fig:llmfailure}B).
Beyond these examples, LLMs often report partially correct or incorrect results, such as identifying only sub paths of anti-patterns or inferring incorrect data flows.}
Despite the growing capabilities and popularity of such assistants, these findings underscore the continued need for human-in-the-loop visual analytics tools like \hooklens, as further discussed in \autoref{sec:LLMEra}.

%% file: REVISION/8_discussion.tex
\section{Discussion}

We discuss the limitations of our study and directions for future work.

\subsection{Scalability and Project Context}
\label{subsec:scalability}
\revised{Our study evaluates \hooklens only on relatively small projects (25--33 components), which limits the generalization of the results to larger-scale projects.
Although \hooklens employs a two-level nested node-link diagram, one participant reports visual clutter when analyzing the study projects (\autoref{sec:qualitative}), suggesting scalability challenges as project size increases.
Therefore, future work should investigate alternative visualization (e.g., treemap, circular tree, and icicle tree)~\cite{SourceCodeComprehension} and interaction techniques to better manage visual complexity in larger projects~\cite{overviewdetail}.
In addition, our evaluation focuses exclusively on unfamiliar projects.
Future studies should examine whether the observed effectiveness of \hooklens generalizes to projects that developers are already familiar with, which more closely reflect real-world development settings.
}

\subsection{Supporting Real-World Workflows for Semantic Validation}
\revised{Although our results suggest that \hooklens effectively supports understanding code structures and detecting anti-patterns that span multiple \components, several refactoring tasks remain challenging.
These include detecting a broader range of anti-patterns, performing fine-grained code-level reasoning within individual \components, reasoning about runtime behavior (e.g., execution timing and data changes), and conducting semantic validation of detected anti-patterns.
Semantic validation is particularly important, as some detected anti-patterns may not represent actual issues without considering runtime context or developer intent.
In this regard, the primary contribution of \hooklens lies in its ability to visually organize analysis results, reducing the cognitive burden of manually tracing control and data flows across \components.
However, our user study shows that most intermediate participants continue to rely on direct source code inspection when performing study tasks, indicating that source code analysis still plays an important role in their workflows and mental models (\autoref{sec:qualitative}).
Therefore, to better support real-world development workflows, \hooklens should be provided as an integrated extension or plugin to conventional code editing tools, enabling not only anti-pattern detection but also semantic validation and other intermediate-level refactoring tasks.
Prior studies show that implementing visualization tools as extensions or plugins to existing development environments is an effective approach for integrating them into established workflows~\cite{LinInterLink, BorowskiGraphBuddy, SourceCodeComprehension, ShatnawiAnalyzingProgram, visdata, GouveiaUsingHTML5}.}

\subsection{Extending Support to the Broader React Ecosystem}
\revised{While \hooklens focuses on the \state and \effect \hooks, which are fundamental to understanding \react applications, extending support to other built-in \hooks and external state management systems remains an important direction for future work.}
As indicated by participants in our user study (\autoref{sec:qualitative}), many developers rely on other built-in \hooks, such as the \texttt{Context} hook~\cite{reacthooks}, as well as external state management systems like \textit{Redux}\footnote{\url{https://redux.js.org/}}.
These mechanisms are also essential for understanding the behavior of \react applications, as they influence rendering processes in ways similar to \state \hooks.
Moreover, they enable data access across \component hierarchies, thereby mitigating anti-patterns such as \textit{prop drilling}.
Nevertheless, extending \hooklens to support these mechanisms introduces two major challenges.

First, the rule-based approach used in \hooklens requires substantial manual effort to expand its coverage, although it produces fast and consistent results.
To address this limitation, LLM-based approaches could offer a promising alternative.
Recent studies demonstrate the effectiveness of LLMs in both syntactic and sematic feature extraction from software~\cite{LLMExtraction1, LLMExtraction2}. 
In our preliminary tests, we also observe that \textit{GPT-5} shows potential for extracting \components, \states, and \effects in a manner comparable to the rule-based extraction logic.

Next, attempting to represent these mechanisms using the current node-link diagram in \hooklens may increase visual complexity by generating excessive nodes and edges or disrupting the layout.
Future work should investigate alternative visualization methods to effectively represent them including other built-in \hooks.
This challenge is common in node-link diagram representations, where various techniques \revised{(e.g., grid layout for grouped graph~\cite{gridgraph} or bubble set~\cite{BubbleSets})} have been explored to address similar issues.

\subsection{Visual Analytics in the AI Era}
\label{sec:LLMEra}

Our study suggests that developers, with the support of \hooklens, achieve a deeper understanding of code structures and detect anti-patterns more effectively than state-of-the-art LLM-based coding assistants (\autoref{sec:llmcompare}). 
However, \revised{our evaluation is conducted with limited prompt engineering and a constrained number of interaction rounds,} and the strong performance of recent models such as \textit{GPT-5} suggests that future coding assistants may increasingly understand broader code contexts and deliver better results.
\revised{In fact, several recent studies and commercial systems propose frameworks and techniques that enable LLMs to retain and reason over richer contextual information~\cite{Codellm, serena}.}
Despite these advances, prior studies~\cite{WeiszAICodeAssistant, MozannarAI-AssistedProgramming, PearceAsleepCopilot} and our findings indicate that LLM-based assistants can still introduce basic errors or overlook subtle but critical dependencies.
Therefore, visual analytics for software remains essential not only for helping developers comprehend complex projects but also for providing a foundation to interpret, monitor, and guide the behavior of LLM-based assistants in emerging coding paradigms.
Future research should thus expand its focus beyond conventional programming environments to emerging ecosystems that integrate LLMs and other AI technologies.
Recent studies such as \textit{NeuroSync}~\cite{NeuroSync} and \textit{DreamGarden}~\cite{EarleDreamGarden} present promising directions by leveraging visualization to support users in understanding and controlling AI-driven code generation processes.

%% file: REVISION/9_conclusion.tex
\section{Conclusion}

\hooklens offers a novel approach to addressing the challenges of understanding \react applications and detecting anti-patterns.
By visually representing the relationships between \components and \hooks, \hooklens enables developers to identify anti-patterns and inspect relative \components and \hooks along with their source code.
Through preliminary interviews and an iterative design process, we develop \hooklens, an interactive visual analytics system for exploring \react applications from the perspective of \hooks and anti-patterns.
Findings from our user study and comparative evaluation show that \hooklens not only supports understanding code structures and detecting anti-patterns, but also highlights that such understanding remains essential even when using LLM-based coding assistants. 
Therefore, extending visual analytics to software frameworks such as \react holds strong potential to help developers build and maintain interactive applications more effectively.

%% file: template.bib
@String{Computing = "Computing" }

@String{Computer = "{IEEE} Computer" }

@String{Academic = "Academic Press" }

@String{Springer = "Springer-Verlag" }

@inproceedings{TurcotteDrAsync,
    author = {Turcotte, Alexi and Shah, Michael D. and Aldrich, Mark W. and Tip, Frank},
    title = {DrAsync: identifying and visualizing anti-patterns in asynchronous JavaScript},
    year = {2022},
    isbn = {9781450392211},
    publisher = {Association for Computing Machinery},
    address = {New York, NY, USA},
    url = {https://doi.org/10.1145/3510003.3510097},
    doi = {10.1145/3510003.3510097},
    booktitle = {Proceedings of the 44th International Conference on Software Engineering},
    pages = {774–785},
    numpages = {12},
    location = {Pittsburgh, Pennsylvania},
    series = {ICSE '22}
}

@INPROCEEDINGS{ShatnawiAnalyzingProgram,
    author={Shatnawi, Anas and Mili, Hafedh and El Boussaidi, Ghizlane and Boubaker, Anis and Guéhéneuc, Yann-Gaël and Moha, Naouel and Privat, Jean and Abdellatif, Manel},
    booktitle={2017 IEEE/ACM 14th International Conference on Mining Software Repositories (MSR)}, 
    title={Analyzing Program Dependencies in Java EE Applications}, 
    year={2017},
    pages={64-74},
    doi={10.1109/MSR.2017.6}
}

@INPROCEEDINGS{ShneidermanTheeyeshaveit,
  author={Shneiderman, B.},
  booktitle={Proceedings 1996 IEEE Symposium on Visual Languages}, 
  title={The eyes have it: a task by data type taxonomy for information visualizations}, 
  year={1996},
  volume={},
  number={},
  pages={336-343},
  doi={10.1109/VL.1996.545307}
}

@INPROCEEDINGS{AlmeidaRustViz,
    author={Almeida, Marcelo and Cole, Grant and Du, Ke and Luo, Gongming and Pan, Shulin and Pan, Yu and Qiu, Kai and Reddy, Vishnu and Zhang, Haochen and Zhu, Yingying and Omar, Cyrus},
    booktitle={2022 IEEE Symposium on Visual Languages and Human-Centric Computing (VL/HCC)}, 
    title={RustViz: Interactively Visualizing Ownership and Borrowing}, 
    year={2022},
    volume={},
    number={},
    pages={1-10},
    doi={10.1109/VL/HCC53370.2022.9833121}
}

@article{FERREIRAPropDrilling,
    title = {Detecting code smells in {React}-based Web apps},
    journal = {Information and Software Technology},
    volume = {155},
    pages = {107111},
    year = {2023},
    issn = {0950-5849},
    doi = {https://doi.org/10.1016/j.infsof.2022.107111},
    url = {https://www.sciencedirect.com/science/article/pii/S0950584922002208},
    author = {Fabio Ferreira and Marco Tulio Valente},
}

@article{FERREIRAMigrateClassToFunction,
title = {Refactoring {React}-based Web apps},
journal = {Journal of Systems and Software},
volume = {215},
pages = {112105},
year = {2024},
issn = {0164-1212},
doi = {https://doi.org/10.1016/j.jss.2024.112105},
url = {https://www.sciencedirect.com/science/article/pii/S016412122400150X},
author = {Fabio Ferreira and Hudson Silva Borges and Marco Tulio Valente},
}

@article{NUNESTScodesmell,
title = {Detection of code smells in react with TypeScript applications},
journal = {Information and Software Technology},
volume = {187},
pages = {107835},
year = {2025},
issn = {0950-5849},
doi = {https://doi.org/10.1016/j.infsof.2025.107835},
url = {https://www.sciencedirect.com/science/article/pii/S0950584925001740},
author = {Maykon Nunes and Carla Bezerra and Fabio Ferreira and Bruno Gois and Marco Tulio Valente},
}

@misc{reactUseState,
  author       = {{Meta Platforms, Inc.}},
  title        = {State: A Component's Memory - React},
  year         = {2025},
  howpublished = {\url{https://react.dev/learn/state-a-components-memory}},
  note         = {Accessed: 2025-09-05}
}

@misc{reactUseEffect,
  author       = {{Meta Platforms, Inc.}},
  title        = {Synchronizing with Effects - React},
  year         = {2025},
  howpublished = {\url{https://react.dev/learn/synchronizing-with-effects}},
  note         = {Accessed: 2025-09-05}
}

@misc{reactRender,
  author       = {{Meta Platforms, Inc.}},
  title        = {Render and Commit - React},
  year         = {2025},
  howpublished = {\url{https://react.dev/learn/render-and-commit}},
  note         = {Accessed: 2025-10-30}
}

@misc{reactLifecycle,
  author       = {{Meta Platforms, Inc.}},
  title        = {Lifecycle of Reactive Effects - React},
  year         = {2025},
  howpublished = {\url{https://react.dev/learn/lifecycle-of-reactive-effects}},
  note         = {Accessed: 2025-10-30}
}

@misc{reacteffect,
  author       = {{Meta Platforms, Inc.}},
  title        = {You Might Not Need an Effect - React},
  year         = {2025},
  howpublished = {\url{https://react.dev/learn/you-might-not-need-an-effect}},
  note         = {Accessed: 2025-01-07}
}

@misc{reactcomponent,
  author       = {{Meta Platforms, Inc.}},
  title        = {Component - React},
  year         = {2025},
  howpublished = {\url{https://react.dev/reference/react/Component}},
  note         = {Accessed: 2025-08-25}
}

@misc{reacthooks,
  author       = {{Meta Platforms, Inc.}},
  title        = {Built-in React Hooks - React},
  year         = {2024},
  howpublished = {\url{https://react.dev/reference/react/hooks}},
  note         = {Accessed: 2025-01-07}
}

@misc{reactcontext,
  author       = {{Meta Platforms, Inc.}},
  title        = {Passing Data Deeply with Context - React},
  year         = {2024},
  howpublished = {\url{https://react.dev/learn/passing-data-deeply-with-context}},
  note         = {Accessed: 2025-01-07}
}

@misc{soPopularTech,
  author = {stackoverflow},
  title = {2024 Developer Survey - Most popular technologies},
  year = {2024},
  howpublished = {\url{https://survey.stackoverflow.co/2024/technology}}
}

@misc{espree,
  author = {{ESLint}},
  title = {Espree},
  year = {2024},
  howpublished = {\url{https://github.com/eslint/js/blob/main/packages/espree/}},
  note         = {Accessed: 2025-10-27}
}

@INPROCEEDINGS{NguyenCodeSmellsWeb,
  author={Nguyen, Hung Viet and Nguyen, Hoan Anh and Nguyen, Tung Thanh and Nguyen, Anh Tuan and Nguyen, Tien N.},
  booktitle={2012 Proceedings of the 27th IEEE/ACM International Conference on Automated Software Engineering}, 
  title={Detection of embedded code smells in dynamic web applications}, 
  year={2012},
  volume={},
  number={},
  pages={282-285},
  doi={10.1145/2351676.2351724}}

@INPROCEEDINGS{FardJavaScriptCodeSmells,
  author={Fard, Amin Milani and Mesbah, Ali},
  booktitle={2013 IEEE 13th International Working Conference on Source Code Analysis and Manipulation (SCAM)}, 
  title={{JSNOSE}: Detecting {JavaScript} Code Smells}, 
  year={2013},
  volume={},
  number={},
  pages={116-125},
  doi={10.1109/SCAM.2013.6648192}}

@misc{papervis,
  author       = {{vdslab}},
  title        = {PaperVis},
  year         = {2022},
  howpublished = {\url{https://github.com/vdslab/papers_vis}},
  note         = {Accessed: commit \texttt{2cc206f}}
}

@INPROCEEDINGS{Confides,
  author={Ha, Sunwoo and Lim, Chaehun and Crouser, R. Jordan and Ottley, Alvitta},
  booktitle={2024 IEEE Visualization and Visual Analytics (VIS)}, 
  title={Confides: A Visual Analytics Solution for Automated Speech Recognition Analysis and Exploration}, 
  year={2024},
  volume={},
  number={},
  pages={271-275},
  doi={10.1109/VIS55277.2024.00062}}

@misc{ConfidesURL,
  author       = {{washuvis}},
  title        = {{Confides: A Visual Analytics Solution for Automated Speech Recognition Analysis and Exploration}},
  year         = {2024},
  howpublished = {\url{https://github.com/washuvis/vis2024confides}},
  note         = {Accessed: commit \texttt{275e235}}
}

@book{SUSStandard,
  author = {Sauro, Jeff and Lewis, James R},
  year = {2016},
  title = {Quantifying the user experience: Practical statistics for user research},
  publisher={Morgan Kaufmann}
}

@ARTICLE{nodelinkdiagram,
    author={Saket, Bahador and Simonetto, Paolo and Kobourov, Stephen and Börner, Katy},
    journal={IEEE Transactions on Visualization and Computer Graphics}, 
    title={Node, Node-Link, and Node-Link-Group Diagrams: An Evaluation}, 
    year={2014},
    volume={20},
    number={12},
    pages={2231-2240},
    doi={10.1109/TVCG.2014.2346422}}

@article{lee2025reacttrace,
author = {Lee, Jay and Ahn, Joongwon and Yi, Kwangkeun},
title = {React-tRace: A Semantics for Understanding React Hooks: An Operational Semantics and a Visualizer for Clarifying React Hooks},
year = {2025},
issue_date = {October 2025},
publisher = {Association for Computing Machinery},
address = {New York, NY, USA},
volume = {9},
number = {OOPSLA2},
url = {https://doi.org/10.1145/3763067},
doi = {10.1145/3763067},
journal = {Proc. ACM Program. Lang.},
month = oct,
articleno = {289},
numpages = {28}
}

@inproceedings{SalvaneschiDebug,
    author = {Salvaneschi, Guido and Mezini, Mira},
    title = {Debugging for reactive programming},
    year = {2016},
    isbn = {9781450339001},
    publisher = {Association for Computing Machinery},
    address = {New York, NY, USA},
    url = {https://doi.org/10.1145/2884781.2884815},
    doi = {10.1145/2884781.2884815},
    booktitle = {Proceedings of the 38th International Conference on Software Engineering},
    pages = {796–807},
    numpages = {12},
    location = {Austin, Texas},
    series = {ICSE '16}
}

@INPROCEEDINGS{BoersmaReactBratus,
    author={Boersma, Stephan and Lungu, Mircea},
    booktitle={2021 Working Conference on Software Visualization (VISSOFT)}, 
    title={React-bratus: Visualising React Component Hierarchies}, 
    year={2021},
    volume={},
    number={},
    pages={130-134},
    doi={10.1109/VISSOFT52517.2021.00025}}

@INPROCEEDINGS{TarnerCodeQuality,
    author={Tarner, Hagen and van den Bongard, Daniel and Beck, Fabian},
    booktitle={2021 Working Conference on Software Visualization (VISSOFT)}, 
    title={Visually Analyzing the Structure and Code Quality of Component-based Web Applications}, 
    year={2021},
    volume={},
    number={},
    pages={160-164},
    doi={10.1109/VISSOFT52517.2021.00031}}

@inproceedings{GuoReactAppScan,
author = {Guo, Zhiyong and Kang, Mingqing and Venkatakrishnan, V.N. and Gjomemo, Rigel and Cao, Yinzhi},
title = {ReactAppScan: Mining React Application Vulnerabilities via Component Graph},
year = {2024},
isbn = {9798400706363},
publisher = {Association for Computing Machinery},
address = {New York, NY, USA},
url = {https://doi.org/10.1145/3658644.3670331},
doi = {10.1145/3658644.3670331},
booktitle = {Proceedings of the 2024 on ACM SIGSAC Conference on Computer and Communications Security},
pages = {585–599},
numpages = {15},
location = {Salt Lake City, UT, USA},
series = {CCS '24}
}

@article{chotisarn2020systematic,
  title={A systematic literature review of modern software visualization},
  author={Chotisarn, Noptanit and Merino, Leonel and Zheng, Xu and Lonapalawong, Supaporn and Zhang, Tianye and Xu, Mingliang and Chen, Wei},
  journal={Journal of Visualization},
  volume={23},
  number={4},
  pages={539--558},
  year={2020},
  publisher={Springer},
  doi = {10.1007/s12650-020-00647-w},
}

@inproceedings{LinInterLink,
author = {Lin, Yanna and Yang, Leni and Li, Haotian and Qu, Huamin and Moritz, Dominik},
title = {InterLink: Linking Text with Code and Output in Computational Notebooks},
year = {2025},
isbn = {9798400713941},
url = {https://doi.org/10.1145/3706598.3714104},
doi = {10.1145/3706598.3714104},
booktitle = {Proceedings of the 2025 CHI Conference on Human Factors in Computing Systems},
articleno = {51},
numpages = {15},
location = {
},
series = {CHI '25}
}

@inproceedings{BankenDebugDataReactive,
author = {Banken, Herman and Meijer, Erik and Gousios, Georgios},
title = {Debugging data flows in reactive programs},
year = {2018},
isbn = {9781450356381},
publisher = {Association for Computing Machinery},
address = {New York, NY, USA},
url = {https://doi.org/10.1145/3180155.3180156},
doi = {10.1145/3180155.3180156},
booktitle = {Proceedings of the 40th International Conference on Software Engineering},
pages = {752–763},
numpages = {12},
location = {Gothenburg, Sweden},
series = {ICSE '18}
}

@INPROCEEDINGS{BorowskiGraphBuddy,
  author={Borowski, Krzysztof and Balis, Bartosz and Orzechowski, Tomasz},
  booktitle={2022 Working Conference on Software Visualization (VISSOFT)}, 
  title={Graph Buddy — an interactive code dependency browsing and visualization tool}, 
  year={2022},
  volume={},
  number={},
  pages={152-156},
  doi={10.1109/VISSOFT55257.2022.00023}}

@INPROCEEDINGS{GouveiaUsingHTML5,
  author={Gouveia, Carlos and Campos, José and Abreu, Rui},
  booktitle={2013 First IEEE Working Conference on Software Visualization (VISSOFT)}, 
  title={Using HTML5 visualizations in software fault localization}, 
  year={2013},
  pages={1-10},
  doi={10.1109/VISSOFT.2013.6650539}}

@inproceedings{DuRapsai,
author = {Du, Ruofei and Li, Na and Jin, Jing and Carney, Michelle and Miles, Scott and Kleiner, Maria and Yuan, Xiuxiu and Zhang, Yinda and Kulkarni, Anuva and Liu, Xingyu and Sabie, Ahmed and Orts-Escolano, Sergio and Kar, Abhishek and Yu, Ping and Iyengar, Ram and Kowdle, Adarsh and Olwal, Alex},
title = {Rapsai: Accelerating Machine Learning Prototyping of Multimedia Applications through Visual Programming},
year = {2023},
isbn = {9781450394215},
url = {https://doi.org/10.1145/3544548.3581338},
doi = {10.1145/3544548.3581338},
booktitle = {Proceedings of the 2023 CHI Conference on Human Factors in Computing Systems},
articleno = {125},
numpages = {23},
location = {Hamburg, Germany},
series = {CHI '23}
}

@inproceedings{EarleDreamGarden,
author = {Earle, Sam and Parajuli, Samyak and Banburski-Fahey, Andrzej},
title = {DreamGarden: A Designer Assistant for Growing Games from a Single Prompt},
year = {2025},
isbn = {9798400713941},
publisher = {Association for Computing Machinery},
address = {New York, NY, USA},
url = {https://doi.org/10.1145/3706598.3714233},
doi = {10.1145/3706598.3714233},
booktitle = {Proceedings of the 2025 CHI Conference on Human Factors in Computing Systems},
articleno = {57},
numpages = {19},
location = {
},
series = {CHI '25}
}

@inproceedings{LuMisty,
author = {Lu, Yuwen and Leung, Alan and Swearngin, Amanda and Nichols, Jeffrey and Barik, Titus},
title = {Misty: UI Prototyping Through Interactive Conceptual Blending},
year = {2025},
isbn = {9798400713941},
url = {https://doi.org/10.1145/3706598.3713924},
doi = {10.1145/3706598.3713924},
booktitle = {Proceedings of the 2025 CHI Conference on Human Factors in Computing Systems},
articleno = {1108},
numpages = {17},
location = {
},
series = {CHI '25}
}

@inproceedings{LiuVAMaintain,
author = {Liu, Kaihua and Reddivari, Sandeep},
title = {Visual Analytics in Software Maintenance: A Systematic Literature Review},
year = {2023},
isbn = {9781450399210},
url = {https://doi.org/10.1145/3564746.3587022},
doi = {10.1145/3564746.3587022},
booktitle = {Proceedings of the 2023 ACM Southeast Conference},
pages = {70–77},
numpages = {8},
location = {Virtual Event, USA},
series = {ACMSE '23}
}

@inproceedings{
jimenez2024swebench,
title={{SWE}-bench: Can Language Models Resolve Real-world Github Issues?},
author={Carlos E Jimenez and John Yang and Alexander Wettig and Shunyu Yao and Kexin Pei and Ofir Press and Karthik R Narasimhan},
booktitle={The Twelfth International Conference on Learning Representations},
year={2024},
url={https://openreview.net/forum?id=VTF8yNQM66}
}

@article{manckinlayHueSaturation,
author = {Mackinlay, Jock},
title = {Automating the design of graphical presentations of relational information},
year = {1986},
issue_date = {April 1986},
publisher = {Association for Computing Machinery},
address = {New York, NY, USA},
volume = {5},
number = {2},
issn = {0730-0301},
url = {https://doi.org/10.1145/22949.22950},
doi = {10.1145/22949.22950},
journal = {ACM Trans. Graph.},
month = apr,
pages = {110–141},
numpages = {32}
}

@inproceedings{TsengCategorical,
author = {Tseng, Chin and Quadri, Ghulam Jilani and Wang, Zeyu and Szafir, Danielle Albers},
title = {Measuring Categorical Perception in Color-Coded Scatterplots},
year = {2023},
isbn = {9781450394215},
publisher = {Association for Computing Machinery},
address = {New York, NY, USA},
url = {https://doi.org/10.1145/3544548.3581416},
doi = {10.1145/3544548.3581416},
booktitle = {Proceedings of the 2023 CHI Conference on Human Factors in Computing Systems},
articleno = {824},
numpages = {14},
location = {Hamburg, Germany},
series = {CHI '23}
}

@book{qiu2024react,
  title={React Anti-Patterns: Build efficient and maintainable React applications with test-driven development and refactoring},
  author={Qiu, Juntao},
  year={2024},
  publisher={Packt Publishing Ltd}
}

@INPROCEEDINGS{SharmaVsAngular,
  author={Sharma, Tarun and Gupta, Shruti and Singh, Uday Raj},
  booktitle={2023 International Conference on Computational Intelligence, Communication Technology and Networking (CICTN)}, 
  title={Analyzing the difference between {ReactJS} and {AngularJS}}, 
  year={2023},
  volume={},
  number={},
  pages={37-42},
  doi={10.1109/CICTN57981.2023.10141276}}

@article{PearceAsleepCopilot,
author = {Pearce, Hammond and Ahmad, Baleegh and Tan, Benjamin and Dolan-Gavitt, Brendan and Karri, Ramesh},
title = {Asleep at the Keyboard? Assessing the Security of GitHub Copilot’s Code Contributions},
year = {2025},
issue_date = {February 2025},
publisher = {Association for Computing Machinery},
address = {New York, NY, USA},
volume = {68},
number = {2},
issn = {0001-0782},
url = {https://doi.org/10.1145/3610721},
doi = {10.1145/3610721},
journal = {Commun. ACM},
month = jan,
pages = {96–105},
numpages = {10}
}

@inproceedings{WeiszAICodeAssistant,
author = {Weisz, Justin D. and Kumar, Shraddha Vijay and Muller, Michael and Browne, Karen-Ellen and Goldberg, Arielle and Heintze, Katrin Ellice and Bajpai, Shagun},
title = {Examining the Use and Impact of an AI Code Assistant on Developer Productivity and Experience in the Enterprise},
year = {2025},
isbn = {9798400713958},
url = {https://doi.org/10.1145/3706599.3706670},
doi = {10.1145/3706599.3706670},
booktitle = {Proceedings of the Extended Abstracts of the CHI Conference on Human Factors in Computing Systems},
articleno = {673},
numpages = {13},
location = {
},
series = {CHI EA '25}
}

@inproceedings{MozannarAI-AssistedProgramming,
author = {Mozannar, Hussein and Bansal, Gagan and Fourney, Adam and Horvitz, Eric},
title = {Reading Between the Lines: Modeling User Behavior and Costs in AI-Assisted Programming},
year = {2024},
isbn = {9798400703300},
url = {https://doi.org/10.1145/3613904.3641936},
doi = {10.1145/3613904.3641936},
booktitle = {Proceedings of the 2024 CHI Conference on Human Factors in Computing Systems},
articleno = {142},
numpages = {16},
location = {Honolulu, HI, USA},
series = {CHI '24}
}

@inproceedings{WuIsmell,
author = {Wu, Di and Mu, Fangwen and Shi, Lin and Guo, Zhaoqiang and Liu, Kui and Zhuang, Weiguang and Zhong, Yuqi and Zhang, Li},
title = {iSMELL: Assembling LLMs with Expert Toolsets for Code Smell Detection and Refactoring},
year = {2024},
isbn = {9798400712487},
url = {https://doi.org/10.1145/3691620.3695508},
doi = {10.1145/3691620.3695508},
booktitle = {Proceedings of the 39th IEEE/ACM International Conference on Automated Software Engineering},
pages = {1345–1357},
numpages = {13},
location = {Sacramento, CA, USA},
series = {ASE '24}
}

@InProceedings{TessaSmell,
author="Tessa, Claudio
and Bochicchio, Matteo
and Fontana, Francesca Arcelli",
editor="Andrikopoulos, Vasilios
and Pautasso, Cesare
and Ali, Nour
and Soldani, Jacopo
and Xu, Xiwei",
title="Exploring Architectural Smells Detection Through LLMs",
booktitle="Software Architecture",
year="2026",
pages="90--98",
isbn="978-3-032-02138-0",
doi={10.1007/978-3-032-02138-0_6}
}

@misc{claudecode,
  author       = {{ANTHROPIC}},
  title        = {Claude Code: Your code's new collaborator},
  year         = {2022},
  howpublished = {\url{https://www.anthropic.com/claude-code}},
  note         = {Accessed: 2025-09-10}
}

@misc{codexcli,
  author       = {{OpenAI}},
  title        = {{Codex CLI: Pair with Codex in your terminal}},
  year         = {2025},
  howpublished = {\url{https://developers.openai.com/codex/cli}},
  note         = {Accessed: 2025-09-10}
}

@misc{geminicli,
  author       = {{Google}},
  title        = {{Gemini CLI: Build, debug \& deploy with AI}},
  year         = {2025},
  howpublished = {\url{https://geminicli.com/}},
  note         = {Accessed: 2026-01-25}
}

@misc{tscompiler,
  author       = {{Microsoft}},
  title        = {Using the Compiler API},
  year         = {2023},
  howpublished = {\url{https://github.com/microsoft/TypeScript/wiki/Using-the-Compiler-API}},
  note         = {Accessed: 2025-10-27}
}

@book{fowler2018refactoring,
  title={Refactoring: improving the design of existing code},
  author={Fowler, Martin},
  year={2018},
  publisher={Addison-Wesley Professional}
}

@article{maintenence,
author = {Khomh, Foutse and Penta, Massimiliano Di and Gu\'{e}h\'{e}neuc, Yann-Ga\"{e}l and Antoniol, Giuliano},
title = {An exploratory study of the impact of antipatterns on class change- and fault-proneness},
year = {2012},
issue_date = {June      2012},
publisher = {Kluwer Academic Publishers},
address = {USA},
volume = {17},
number = {3},
issn = {1382-3256},
url = {https://doi.org/10.1007/s10664-011-9171-y},
doi = {10.1007/s10664-011-9171-y},
journal = {Empirical Softw. Engg.},
month = jun,
pages = {243–275},
numpages = {33}
}

@inproceedings{AST1,
author = {Visser, Eelco},
title = {Meta-programming with Concrete Object Syntax},
year = {2002},
isbn = {3540442847},
publisher = {Springer-Verlag},
address = {Berlin, Heidelberg},
booktitle = {Proceedings of the 1st ACM SIGPLAN/SIGSOFT Conference on Generative Programming and Component Engineering},
pages = {299–315},
numpages = {17},
series = {GPCE '02},
doi = {10.1007/3-540-45821-2_19}
}

@ARTICLE{gridgraph,
  author={Yoghourdjian, Vahan and Dwyer, Tim and Gange, Graeme and Kieffer, Steve and Klein, Karsten and Marriott, Kim},
  journal={IEEE Transactions on Visualization and Computer Graphics}, 
  title={High-Quality Ultra-Compact Grid Layout of Grouped Networks}, 
  year={2016},
  volume={22},
  number={1},
  pages={339-348},
  doi={10.1109/TVCG.2015.2467251}}

@INPROCEEDINGS{overviewdetail,
  author={Han, Chang and Lieffers, Justin and Morrison, Clayton and Isaacs, Katherine E.},
  booktitle={2024 IEEE Visualization and Visual Analytics (VIS)}, 
  title={An Overview+Detail Layout for Visualizing Compound Graphs}, 
  year={2024},
  volume={},
  number={},
  pages={136-140},
  doi={10.1109/VIS55277.2024.00035}}

@INPROCEEDINGS{viscall,
  author={LaToza, Thomas D. and Myers, Brad A.},
  booktitle={2011 IEEE Symposium on Visual Languages and Human-Centric Computing (VL/HCC)}, 
  title={Visualizing call graphs}, 
  year={2011},
  volume={},
  number={},
  pages={117-124},
  doi={10.1109/VLHCC.2011.6070388}}

@INPROCEEDINGS{visdata,
  author={Ishio, Takashi and Etsuda, Shogo and Inoue, Katsuro},
  booktitle={2012 20th IEEE International Conference on Program Comprehension (ICPC)}, 
  title={A lightweight visualization of interprocedural data-flow paths for source code reading}, 
  year={2012},
  volume={},
  number={},
  pages={37-46},
  doi={10.1109/ICPC.2012.6240506}}

@inproceedings{NeuroSync,
author = {Zhang, Wenshuo and Shen, Leixian and Xu, Shuchang and Wang, Jindu and Zhao, Jian and Qu, Huamin and Yuan, Lin-Ping},
title = {NeuroSync: Intent-Aware Code-Based Problem Solving via Direct LLM Understanding Modification},
year = {2025},
isbn = {9798400720376},
url = {https://doi.org/10.1145/3746059.3747668},
doi = {10.1145/3746059.3747668},
booktitle = {Proceedings of the 38th Annual ACM Symposium on User Interface Software and Technology},
articleno = {30},
numpages = {19},
location = {
},
series = {UIST '25}
}

@inproceedings{DetectingCodesmellsGPT,
author = {Silva, Luciana Lourdes and Silva, Janio Rosa da and Montandon, Joao Eduardo and Andrade, Marcus and Valente, Marco Tulio},
title = {Detecting Code Smells using ChatGPT: Initial Insights},
year = {2024},
isbn = {9798400710476},
url = {https://doi.org/10.1145/3674805.3690742},
doi = {10.1145/3674805.3690742},
booktitle = {Proceedings of the 18th ACM/IEEE International Symposium on Empirical Software Engineering and Measurement},
pages = {400–406},
numpages = {7},
location = {Barcelona, Spain},
series = {ESEM '24}
}

@INPROCEEDINGS{SourceCodeComprehension,
  author={Bacher, Ivan and Namee, Brian Mac and Kelleher, John D.},
  booktitle={2016 IEEE Working Conference on Software Visualization (VISSOFT)}, 
  title={On Using Tree Visualisation Techniques to Support Source Code Comprehension}, 
  year={2016},
  volume={},
  number={},
  pages={91-95},
  doi={10.1109/VISSOFT.2016.8}}

@INPROCEEDINGS{LLMExtraction1,
  author={Talukder, A A Talha and Alam, Omar and Azim, Akramul},
  booktitle={2025 IEEE International Conference on Information Reuse and Integration and Data Science (IRI)}, 
  title={Leveraging LLMs for Automatic Feature Extraction in Embedded Systems to Support Software Reuse}, 
  year={2025},
  volume={},
  number={},
  pages={1-6},
  doi={10.1109/IRI66576.2025.00009}}

@INPROCEEDINGS{LLMExtraction2,
  author={Ignatyev, V. N. and Shimchik, N. V. and Panov, D. D. and Mitrofanov, A. A.},
  booktitle={2024 Ivannikov Memorial Workshop (IVMEM)}, 
  title={Large language models in source code static analysis}, 
  year={2024},
  volume={},
  number={},
  pages={28-35},
  doi={10.1109/IVMEM63006.2024.10659715}}

@misc{serena,
  author       = {{Oraios AI}},
  title        = {About Serena},
  year         = {2025},
  howpublished = {\url{https://oraios.github.io/serena/01-about/000_intro.html}},
  note         = {Accessed: 2026-01-04}
}

@inproceedings{Codellm,
author = {Krishna, Rahul and Pan, Rangeet and Sinha, Saurabh and Tamilselvam, Srikanth and Pavuluri, Raju and Vukovic, Maja},
title = {Codellm-Devkit: A Framework for Contextualizing Code LLMs with Program Analysis Insights},
year = {2025},
isbn = {9798400712760},
url = {https://doi.org/10.1145/3696630.3728555},
doi = {10.1145/3696630.3728555},
booktitle = {Proceedings of the 33rd ACM International Conference on the Foundations of Software Engineering},
pages = {308–318},
numpages = {11},
location = {Clarion Hotel Trondheim, Trondheim, Norway},
}

@ARTICLE{BubbleSets,
  author={Collins, Christopher and Penn, Gerald and Carpendale, Sheelagh},
  journal={IEEE Transactions on Visualization and Computer Graphics}, 
  title={Bubble Sets: Revealing Set Relations with Isocontours over Existing Visualizations}, 
  year={2009},
  volume={15},
  number={6},
  pages={1009-1016},
  doi={10.1109/TVCG.2009.122}}

@ARTICLE{kim21tvcg,
  author={Kim, Youngtaek and Kim, Jaeyoung and Jeon, Hyeon and Kim, Young-Ho and Song, Hyunjoo and Kim, Bohyoung and Seo, Jinwook},
  journal={IEEE Transactions on Visualization and Computer Graphics}, 
  title={Githru: Visual Analytics for Understanding Software Development History Through Git Metadata Analysis}, 
  year={2021},
  volume={27},
  number={2},
  pages={656-666},
  doi={10.1109/TVCG.2020.3030414}}

@INPROCEEDINGS{kim21pvis,
  author={Kim, Youngtaek and Jeon, Hyeon and Kim, Young-Ho and Ki, Yuhoon and Song, Hyunjoo and Seo, Jinwook},
  booktitle={2021 IEEE 14th Pacific Visualization Symposium (PacificVis)}, 
  title={Visualization Support for Multi-criteria Decision Making in Software Issue Propagation}, 
  year={2021},
  volume={},
  number={},
  pages={81-85},
  doi={10.1109/PacificVis52677.2021.00018}}

@misc{sparkGenUI,
  author    = {Park, Seokhyeon and Lee, Soohyun and Choi, Eugene and Kim, Hyunwoo and Kweon, Minkyu and Song, Yumin and Seo, Jinwook},
  title     = {Bridging Gulfs in UI Generation through Semantic Guidance},
  year      = {2026},
  archivePrefix = {arXiv},
  eprint    = {2601.19171},
  primaryClass={cs.HC},
  doi =       {10.48550/arXiv.2601.19171},
    url={https://arxiv.org/abs/2601.19171}, 
}
